\documentclass[10pt,aps,prc,floatfix,twocolumn,nofootinbib,superscriptaddress]{revtex4-1}
\usepackage{graphicx,amsmath,amssymb,bm}
\usepackage{amsfonts}

\usepackage[utf8]{inputenc}
\usepackage{verbatim}
\usepackage{float}
\usepackage[labelfont={small},font=small,subrefformat=parens,caption=false]{subfig}
\usepackage{cancel}
\usepackage{multirow}
\usepackage{array}
\usepackage{xparse}
\usepackage{xspace}
\usepackage{bigstrut}
\usepackage{physics}
\usepackage{xcolor}
\usepackage{soulpos}  
\usepackage{dsfont}

\usepackage{dcolumn}
\newcolumntype{d}[1]{D{.}{.}{#1}}
\usepackage{xurl}
\usepackage[pdfencoding=auto, pdfpagelabels]{hyperref}

\usepackage[overload]{textcase}

\definecolor{linkcolor}{rgb}{0,0,0.40} 
\hypersetup{%
    pdfsubject=Paper,
    pdfkeywords={nuclear physics} {Bayesian} {chiral EFT} {perturbative QCD}{equation of state},
    unicode = true,
    breaklinks = true,
    colorlinks = true,
    linkcolor = linkcolor,
    citecolor = linkcolor,
    menucolor = linkcolor,
    urlcolor = linkcolor
}

\usepackage{cellspace}
\graphicspath{{./paper_figures/}}

\makeatletter
\newcommand\newsubcommand[3]{\newcommand#1{#2\sc@sub{#3}}}
\def\sc@sub#1{\def\sc@thesub{#1}\@ifnextchar_{\sc@mergesubs}{_{\sc@thesub}}}
\def\sc@mergesubs_#1{_{\sc@thesub#1}}

\newcommand\newsupcommand[3]{\newcommand#1{#2\sc@sup{#3}}}
\def\sc@sup#1{\def\sc@thesup{#1}\@ifnextchar^{\sc@mergesups}{^{\sc@thesup}}}
\def\sc@mergesups^#1{^{\sc@thesup#1}}
\makeatother

\DeclareMathAlphabet{\mathbcal}{OMS}{cmsy}{b}{n}

\newcommand{\cbar}{\bar c}

\newcommand{\fmiq}{\, \text{fm}^{-3}}

\newcommand{\ordervec}{\vec}

\newcommand{\inputvec}{\mathbf}

\newsubcommand{\ckvec}{\ordervec{c}}{k}

\newsubcommand{\bkvec}{\ordervec{b}}{k}

\newsubcommand{\ckvecset}{\ordervec{\inputvec{c}}}{k}

\newsubcommand{\ckvecapprox}{\mathbf{c}'}{k}
\newsubcommand{\ckvecapproxset}{\mathbf{C}'}{k}

\newsubcommand{\bkvecapprox}{\mathbf{b}'}{k}
\newsubcommand{\bkvecset}{\mathbf{B}}{k}
\newsubcommand{\bkvecapproxset}{\mathbf{B}'}{k}

\newcommand{\genobs}{y}

\newsubcommand{\genobsvec}{\ordervec{\genobs}}{k}
\newsubcommand{\genobsvecset}{\ordervec{\inputvec{\genobs}}}{k}

\newsubcommand{\akvec}{\mathbf{a}}{k}

\newsubcommand{\akvecapprox}{\mathbf{a}'}{k}
\newsubcommand{\akvecset}{\mathbf{A}}{k}
\newsubcommand{\akvecapproxset}{\mathbf{A}'}{k}

{}  

\newcommand{\given}{\,|\,}  

\def\diffd{\mathrm{d}}  

\DeclareDocumentCommand\differential{ o g d() }{ 
    \IfNoValueTF{#2}{
        \IfNoValueTF{#3}
            {\diffd\IfNoValueTF{#1}{}{^{#1}}}
            {\mathinner{\diffd\IfNoValueTF{#1}{}{^{#1}}\argopen(#3\argclose)}}
        }
        {\mathinner{\diffd\IfNoValueTF{#1}{}{^{#1}}#2} \IfNoValueTF{#3}{}{(#3)}}
    }

\newcommand{\pathd}{\mathcal{D}}  

\DeclareDocumentCommand\pathdifferential{ o g d() }{ 
    \IfNoValueTF{#2}{
        \IfNoValueTF{#3}
            {\pathd\IfNoValueTF{#1}{}{^{#1}}}
            {\mathinner{\pathd\IfNoValueTF{#1}{}{^{#1}}\argopen(#3\argclose)}}
        }
        {\mathinner{\pathd\IfNoValueTF{#1}{}{^{#1}}#2} \IfNoValueTF{#3}{}{(#3)}}
    }

\newcommand{\ChiEFT}{$\chi$EFT}
\newcommand{\GeV}{\, \text{GeV}}
\newcommand{\Lambdabar}{\bar{\Lambda}}

\definecolor{lightblue}{rgb}{.87,.95,.99}

\newcommand{\orcid}[1]{\href{https://orcid.org/#1}{\includegraphics[scale=0.055]{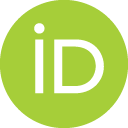}}}

\begin{document}

\title{Microscopic constraints for the equation of state and structure of neutron stars:\\a Bayesian model mixing framework}

\author{A.~C. Semposki
\orcid{0000-0003-2354-1523}}
\email{as727414@ohio.edu}
\affiliation{Department of Physics and Astronomy and Institute of Nuclear and Particle Physics, Ohio University, Athens, OH 45701, USA}

\author{C. Drischler
\orcid{0000-0003-1534-6285}}
\email{drischler@ohio.edu}
\affiliation{Department of Physics and Astronomy and Institute of Nuclear and Particle Physics, Ohio University, Athens, OH 45701, USA}
\affiliation{Facility for Rare Isotope Beams, Michigan State University, East Lansing, MI 48824, USA}

\author{R.~J. Furnstahl 
\orcid{0000-0002-3483-333X}}
\email{furnstahl.1@osu.edu}
\affiliation{Department of Physics, The Ohio State University, Columbus, OH 43210, USA}

\author{D.~R. Phillips
\orcid{0000-0003-1596-9087}}
\email{phillid1@ohio.edu}
\affiliation{Department of Physics and Astronomy and Institute of Nuclear and Particle Physics, Ohio University, Athens, OH 45701, USA}
\affiliation{Department of Physics, Chalmers University of Technology, SE-41296 G\"oteborg, Sweden}

\date{\today}

\begin{abstract}
Bayesian model mixing (BMM) is a statistical technique that can combine constraints from different regions of an input space in a principled way. Here we extend our BMM framework for the equation of state (EOS) of strongly interacting matter from symmetric nuclear matter to asymmetric matter, specifically focusing on zero-temperature, charge-neutral, $\beta$-equilibrated matter. 
We use Gaussian processes (GPs) to infer constraints on the neutron star matter EOS at intermediate densities from two different microscopic theories: chiral effective field theory ($\chi$EFT) at baryon densities around nuclear saturation, $n_B \sim n_0$, and perturbative QCD at asymptotically high baryon densities, $n_B \geqslant 20 n_0$. The uncertainties of the $\chi$EFT and pQCD EOSs are obtained using the BUQEYE truncation error model. We demonstrate the flexibility of our framework through the use of two categories of GP kernels: conventional stationary kernels and a non-stationary changepoint kernel.
We use the latter to explore potential constraints on the dense matter EOS by including exogenous data representing theory predictions and heavy-ion collision measurements at densities $\geqslant 2n_0$. We also use our EOSs 
to obtain neutron star mass-radius relations and their uncertainties. Our framework, whose implementation will be available through a GitHub repository, provides a prior distribution for the EOS that can be used in large-scale neutron-star inference frameworks.
\end{abstract}

\maketitle


\section{Motivation} \label{sec:intro}

Understanding strongly interacting matter across the wide range of densities within neutron stars is a challenging task. Quantum Chromodynamics (QCD) describes strongly interacting matter at all densities, but its non-perturbative nature at low energies and densities makes it computationally intractable in this region of the phase diagram.
This difficulty is addressed by effective descriptions of QCD, i.e., effective field theories that enable order-by-order calculations of many-body forces present in nuclear matter~\cite{Hammer:2012id, Drischler:2021kxf, Machleidt:2016rvv}. At densities below twice the nuclear saturation density ($n_{0}\approx 0.164$ fm$^{-3}$~\cite{Drischler:2024ebw}), chiral effective field theory ($\chi$EFT) with Weinberg power counting is the state-of-the-art approach to describing neutron-rich matter~\cite{ Drischler:2021kxf, Chatziioannou:2024tjq, Tews:2020hgp}. At very high densities, above $20n_0$--$40n_0$, QCD becomes perturbative in the strong coupling constant, and perturbative QCD (pQCD) can be used to describe cold quark matter~\cite{Kurkela:2009gj, Gorda:2018gpy, Gorda:2021znl, Gorda:2023mkk}.

Neither theory can reliably describe intermediate densities, such as those present in the inner cores of neutron stars. Developing a numerically feasible theoretical description across this intermediate region between $\chi$EFT and pQCD, with controlled and fully quantified uncertainties, is extremely difficult. Complications include the potential existence of exotic states of matter, e.g., meson condensates, and phase transitions~\cite{Chatziioannou:2015uea, Lattimer:2021emm, Han:2019bub, Somasundaram:2021clp, Lattimer:2004pg}. Bayesian methods can aid in the formulation of different equations of state (EOS) that arise from incorporating these characteristics of strongly interacting matter across a wide range of densities~\cite{Komoltsev:2021jzg, Gorda:2022jvk, Gorda:2023usm, Komoltsev:2023zor, Koehn:2024set}. They thereby enable a unified description of dense matter with rigorous uncertainty quantification (UQ)~\cite{Semposki:2024vnp, MUSES:2023hyz}.

Bayesian model mixing (BMM) combines probability distributions obtained from multiple models across a chosen input space in a principled manner.
In this paper, we adapt our BMM framework, introduced in Ref.~\cite{Semposki:2024vnp} (which we henceforth denote as Paper~I), to neutron star matter (Sec.~\ref{sec:bmm}). As in Paper~I, we use a Gaussian process (GP) to carry out this mixing. GPs can encode the desired physics of the system through the choice of a covariance function (or \textit{kernel}) and a mean function, thereby imposing a prior on the underlying EOS function space. Our mixed-model GP EOS is trained using data from the \ChiEFT\ and pQCD EOSs, which are explained in Sec.~\ref{sec:EOSs}. The GP EOS has quantified uncertainties that are conditional on the model-space prior.  Our framework is modular and flexible, and can easily be used with GP priors beyond the several that we explore here.

In Paper~I we developed this framework for the case of symmetric nuclear matter (SNM). Here, we turn our attention to neutron star matter: strongly-interacting matter that is charge neutral and in $\beta$-equilibrium. Paper~I also only explored a single GP kernel to describe the entire density region from $\chi$EFT through pQCD---the widely-used stationary squared-exponential radial basis function kernel. 
In this paper, we explore other stationary kernel choices (Sec.~\ref{sec:stationaryGPs}). We also develop a non-stationary changepoint kernel that allows for different kernels in different regions of density (Sec.~\ref{sec:changepointGPs}). This compound kernel mitigates concerns associated with trying to describe such a large density range with a density-independent kernel.

We update the changepoint kernel with exogenous data in the intermediate density region, i.e., information on the EOS that is not included in the $\chi$EFT and pQCD constraints. Such information could be from potential heavy-ion collision measurements or model predictions. We show how the process of updating the changepoint kernel in light of such data is carried out, and elucidate the implications such measurements may have for the EOS of dense matter. We propagate the resulting EOSs from both kernel types, along with their uncertainties, to the mass-radius relation for static neutron stars by solving the Tolman-Oppenheimer-Volkoff equations~\cite{tolman1939, oppenheimer1939} in Secs.~\ref{sec:stationaryGPs} and~\ref{sec:changepointGPs}. This allows us to compare the $M-R$ posterior found under different kernel assumptions to current NICER and LIGO-Virgo results and also to explore the impact of potential future exogenous data on neutron star mass-radius constraints. In Sec.~\ref{sec:summary} we summarize and discuss how our framework could interface with future neutron star and multi-messenger observations.

We provide access to our framework through a GitHub repository~\cite{EOS_BMM_ANM}. Throughout this work, we use natural units, i.e., $\hbar = c = 1$.


\section{Equations of state} \label{sec:EOSs}

In this section, we discuss the EOSs we use at low ($\chi$EFT) and high (pQCD) densities, and the quantification of their truncation errors through the BUQEYE model~\cite{BUQEYEsoftware, Melendez:2019izc, Drischler:2020yad}. We stress that our BMM framework is agnostic to the choice of EOS and, hence, the details of the Hamiltonian and many-body method used to compute the properties of low-density nuclear matter. However, we do assume that we possess an EOS with quantified model uncertainties.

\subsection{Dense matter from chiral EFT (\texorpdfstring{$\chi$EFT}{chiral EFT})} \label{sec:chiralEFT}

Chiral effective field theory is the state-of-the-art description of QCD at low densities (i.e., $\leqslant 2n_{0}$). To construct the EOS for charge-neutral, $\beta$-equilibrated matter, we use $\chi$EFT predictions for the energy per particle $E(n_B)/A$ as a function of the baryon number density $n_B$, through N$^3$LO in the $\chi$EFT expansion, for SNM and pure neutron matter (PNM). These are calculated using fourth-order many-body perturbation theory (MBPT)~\cite{Drischler:2017wtt, Drischler:2020yad, Drischler:2020hwi} and the nonlocal chiral $NN$ interactions developed by Entem, Machleidt, and Nosyk~\cite{Entem:2015xwa}, together with $3N$ forces at the same momentum cutoff and $\chi$EFT order. The corresponding $3N$ low-energy constants (LECs), $c_{D}$ and $c_{E}$, were fit to the triton binding energy and adjusted to the empirical nuclear saturation point in SNM~\cite{Drischler:2017wtt}. 

\begin{figure}[t]
    \centering
    \includegraphics[width=\columnwidth]{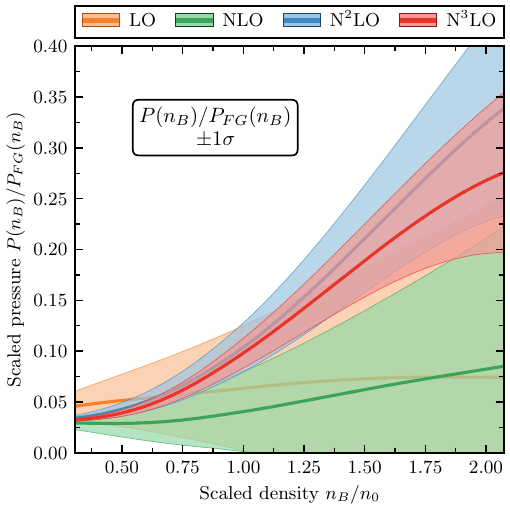}
    \caption{Order-by-order results for the pressure $P(n_B)$ of charge-neutral, $\beta$-equilibrated matter from $\chi$EFT, including leptonic contributions, scaled using the Fermi gas pressure $P_{FG}(n_B)$. The 1$\sigma$ uncertainty bands shown are calculated using the BUQEYE truncation error model~\cite{Melendez:2019izc}. The use of the full covariance matrices of Refs.~\cite{Drischler:2020hwi, Drischler:2020yad} means that the full inter-density correlations are included in this EOS.}
    \label{fig:cheft_anm}
\end{figure}

To determine the neutron-star matter (NSM) EOS from these predictions, we follow the calculation of Ref.~\cite{Drischler:2020fvz} closely. The total energy per baryon of charge-neutral, $\beta$-equilibrated NSM can be determined by expanding the baryon energy per particle about SNM ($\delta = 0$) in terms of the isospin asymmetry parameter $\delta=1-2x$ and truncating after the leading term, which is quadratic in $\delta$.\footnote{
We emphasize that our Bayesian model mixing framework is agnostic about how the beta-equilibrated EOS is obtained. Explicit microscopic calculations of the neutron star EOS could be used instead of an approximate treatment which truncates the expansion in  powers of the isospin asymmetry.
} (Here, $x=n_p/n_B$ is the proton fraction, and $n_p$ is the proton number density.) Explicit microscopic calculations of isospin asymmetric matter based on chiral $NN$ and $3N$ forces indicate that this is a reasonable approximation at least up to $n_0$~\cite{Drischler:2013iza,Drischler:2015eba,Wen:2020nqs}.
This yields, in terms of the energy density in SNM 
and PNM with the rest-mass contribution, and the electron and muon energy densities~\cite{Drischler:2020fvz},
\begin{align}
    \label{eq:nsm_energy}
    \varepsilon_{\textrm{NSM}}(n_B,x,n_e,n_\mu) &= (1-2x)^{2}\varepsilon_{\textrm{PNM}}(n_B) 
        \notag \\
    & \quad \null 
    + 4x(1-x)\varepsilon_{\textrm{SNM}}(n_B) \notag \\
    &\quad+ \varepsilon_{e}(n_e) + \varepsilon_{\mu}(n_\mu).
\end{align} 
The corresponding energy per baryon is then obtained by $E_{\textrm{NSM}} =\varepsilon_{\textrm{NSM}} / n_B$, and similarly for the energies per baryon in PNM and SNM.

The energy density of an ultra-relativistic, degenerate electron gas is given by 
\begin{equation}
    \label{eq:electron_energy}
    \varepsilon_{e} = \frac{k_{F,e}^4}{4\pi^{2}},
\end{equation}
and the energy density of a relativistic muon gas is
\begin{align}
    \varepsilon_{\mu} &= \frac{1}{8\pi^{2}} \Bigg[ k_{F,\mu} \sqrt{k_{F,\mu}^{2} + m_{\mu}^{2}}(2k_{F,\mu}^{2} + m_{\mu}^{2}) \notag \\
    &\quad- m_{\mu}^{4} \ln{\left( \frac{k_{F,\mu} + \sqrt{k_{F,\mu}^{2} + m_{\mu}^{2}}}{m_{\mu}} \right)} \Bigg],
\end{align}
where the leptonic Fermi momenta are $k_{F,\ell} = \sqrt[3]{3\pi^{2}n_{B}x_{\ell}}$, with $x_\ell$ being the electron or muon fraction, and $m_{\ell}$ is the corresponding lepton mass. Muons only begin appearing in the system at densities for which $\mu_{e} \geqslant m_{\mu}$, and so it becomes energetically favorable to achieve charge neutrality through electrons and muons, rather than just electrons. 

We also calculate the (density-dependent) nuclear symmetry energy, $E_\textrm{sym}$, as
\begin{equation}
    \label{eq:sym_energy}
    E_{\textrm{sym}}(n_B) = E_{\textrm{PNM}}(n_B) - E_{\textrm{SNM}}(n_B).
\end{equation}
We then enforce the $\beta$-equilibrium and charge-neutrality condition of NSM at fixed density by invoking
\begin{subequations}\label{eq:betaequilibrium}
\begin{align}
    \mu_{n} - \mu_{p} &= 4(1-2x)E_{\textrm{sym}}(n_B) = \mu_{e} = \mu_{\mu} ,\\
     n_p& =n_e + n_\mu.
    \end{align}
\end{subequations}   
Here, $\mu_{j} = \partial \varepsilon_j / \partial n_j$ corresponds to the chemical potential of a given particle species $j$. 
From Eqs.~\eqref{eq:electron_energy}--\eqref{eq:betaequilibrium}, we can determine the proton fraction $x(n_B)$ corresponding to charge-neutral, $\beta$-equilibrated matter at a particular density, and hence evaluate the total energy per baryon from Eq.~\eqref{eq:nsm_energy}. The proton fraction is calculated using the mean values of the random variables $E_{\textrm{PNM}}$, $E_{\textrm{SNM}}$, and $E_{\textrm{sym}}$. Hence it is a deterministic, not a random, variable. 
$E_{e}$ and $E_{\mu}$ are also not treated as random variables, so the lepton energies do not contribute to the UQ that follows.
The $\beta$-equilibrium proton fraction is then $< 8$\% for $n_B \leqslant 2n_0$, consistent with the findings in Ref.~\cite{Drischler:2020fvz}. Treating $x(n_B)$ as a random variable would allow us to propagate uncertainties in it induced by the PNM and SNM equations of state through to the neutron-star matter EOS. However, the EOS of neutron-star matter is quite close to that of pure-neutron matter, so these additional uncertainties are small compared to other entries in the NSM EOS uncertainty budget.

We use the results of the $\chi$EFT calculations presented in Refs.~\cite{Drischler:2020yad,Drischler:2020hwi} for the energy per baryon of SNM, PNM, and the related symmetry energy. Each observable is described by a GP that uses a squared exponential radial basis function (RBF) kernel; this kernel is described by a correlation ``length'' and marginal variance. (See Sec.~\ref{sec:stationarykernels} for a brief presentation of GPs and Refs.~\cite{Melendez:2019izc, Semposki:2024vnp} for more extensive discussions and applications to nuclear physics.) To account for correlations between the observables $E_{\textrm{PNM}}$ and $E_{\textrm{SNM}}$, Ref.~\cite{Drischler:2020yad} used \textit{multi-task GPs}, in which a covariance function that describes both inter-density and inter-observable correlations is constructed by combining the length scales of the two observables' individual GP descriptions.
From the provided Jupyter notebooks~\cite{BUQEYEsoftware}, we extract the means, standard deviations, and full covariance matrices of $E_{\textrm{SNM}}$, $E_{\textrm{PNM}}$, and $E_{\textrm{sym}}$. These results contain the correlated BUQEYE EFT truncation errors~\cite{Melendez:2019izc}. For a complete description of how these results are obtained using the BUQEYE truncation error model, see Secs.~II and~III of Ref.~\cite{Drischler:2020yad}.%
\footnote{We note that Ref.~\cite{Drischler:2020yad} constructed the GP descriptions of $E_{\textrm{PNM}}$, $E_{\textrm{SNM}}$, and $E_{\textrm{sym}}$ as a function of the Fermi momentum, $k_{F}$. We extract covariance matrices for these observables here as a function of the baryon number density $n_B$.}

To compute the covariance function of neutron-star matter we need the cross-covariance function between the energy per baryon of PNM and that of SNM. 
We obtain this from the covariance function of the symmetry energy~\eqref{eq:sym_energy} by writing:
\begin{align}
    \label{eq:symenergycov}
    \kappa_{E_\textrm{sym}}(n_B, n_B') &=\kappa_{E_\textrm{PNM}}(n_B, n_B') + \kappa_{E_\textrm{SNM}}(n_B, n_B') \nonumber \\ &\quad- 2 \rho \kappa_{E_\textrm{corr}}(n_B, n_B').
\end{align}
Here $\kappa_{E_\textrm{PNM}}$ and $\kappa_{E_\textrm{SNM}}$ are the corresponding covariance functions for the energy per baryon of PNM and SNM, and $\rho \kappa_{E_\textrm{corr}}$ is the combined correlation coefficient and covariance matrix between the two types of matter. Equation~\eqref{eq:symenergycov} then allows us to numerically extract the desired cross-covariance function, $\rho \kappa_{E_\textrm{corr}}(n_B, n_B')$.

The covariance function of neutron star matter's energy per baryon is then constructed from Eq.~(\ref{eq:nsm_energy}). 
\begin{equation}
\label{eq:nsm_cov_energy}
\begin{split}
        \kappa_{E_{\textrm{NSM}}}&(n_B,n_B') \\ &= (1-2x)^{2} \kappa_{E_{\textrm{PNM}}}(n_B,n_B')(1-2x')^{2} \\ &\quad+ 4x (1-x) \kappa_{E_{\textrm{SNM}}}(n_B,n_B')4x'(1-x')  \\ 
    &\quad+ \rho \kappa_{E_{\textrm{corr}}}(n_B,n_B') \bigl[ (1-2x)^{2}4x'(1-x') \\ &\quad+ (1-2x')^{2}4x(1-x) \bigr].
\end{split}
\end{equation}
Here, $x \equiv x(n_B)$, $x' \equiv x(n_B')$.\footnote{When forming the covariance \textit{matrix}, $n_{B}, n_{B}' \rightarrow \vec{n}_{B}, \vec{n}_{B}'$ and we take outer products of the resulting vector terms.} The diagonal terms of the covariance function \eqref{eq:nsm_cov_energy}, where $n_B'=n_B$, reduce to the standard sum in quadrature of two Gaussian random variables with correlation coefficient $\rho$.

We calculate the pressure, $P(n_B)$, of neutron star matter by employing the usual thermodynamic relation
\begin{equation} \label{eq:pressure_nsm}
    P(n_B) = n_B^{2} \dv{n_B} E_{\textrm{NSM}}(n_B).
\end{equation}
The mean values are thus obtained from the mean NSM energies via differentiation of Eq.~\eqref{eq:nsm_energy}. The covariance function of the pressure is computed numerically from Eq.~\eqref{eq:nsm_cov_energy}, which involves taking mixed partial derivatives with respect to the density (see Eq.~\eqref{eq:nsm_cov_pressure} and the surrounding text for details). The resulting $P(n_B)$ mean 
and standard deviation (square root of the diagonal elements of the pressure covariance matrix) are shown in Fig.~\ref{fig:cheft_anm}. 

We also attach the low-density EOS from Refs.~\cite{Negele:1972zp, Baym:1971pw} to the $\chi$EFT EOS at the typical crust-core transition density of $n_B = 0.5n_{0}$. Together, the crust and $\chi$EFT EOSs cover all baryon densities $\leqslant 2n_{0}$.


\subsection{Dense matter from perturbative QCD} \label{sec:pQCD}

We use the EOS from pQCD at high densities.
We do not know the intermediate densities at which non-perturbative effects, e.g., pairing and renormalons, emerge, so we choose two densities, $20n_0$ and $40n_0$, as lower limits on the validity of pQCD in our analysis.

In the calculations of Ref.~\cite{Gorda:2023mkk} the pressure is given as a function of the quark chemical potential $\mu_{q}$, through an expansion in the strong coupling $\alpha_{s}$~\cite{Gorda:2023mkk}:
\begin{align}
    \label{eq:pressurepqcd2023}
    \frac{P(\mu_q)}{P_{FG}(\mu_q)} &\simeq 1 + a_{1,1} \left(\frac{\alpha_{s}(\Lambdabar)}{\pi}\right) \nonumber \\ &\quad+ 3\left(\frac{\alpha_{s}(\Lambdabar)}{\pi}\right)^{2} \bigg[ a_{2,1} \ln\left(\frac{3\alpha_{s}(\Lambdabar)}{\pi}\right) \nonumber \\ 
    &\quad+ a_{2,2}\ln\frac{\bar{\Lambda}}{2\mu_q} + a_{2,3}\bigg] + \mathcal{O}(\alpha_{s}^{3}),
\end{align}
with coefficients $a_{1,1} = -2$, $a_{2,1} = -1$, $a_{2,2} = -3$, and $a_{2,3} = -5.0021$ for cold, three-flavor ($N_f=3$) $\beta$-equilibrated quark matter. For full details of this expression, we refer the reader to Refs.~\cite{Kurkela:2009gj, Gorda:2023mkk}. In Eq.~\eqref{eq:pressurepqcd2023}, $\Lambdabar = 2\mu_q$, and the relativistic Fermi gas (FG) pressure is 
\begin{equation}
    \label{eq:pqcdfgpressure}
    P_{FG}(\mu_q) = \frac{3\mu_q^{4}}{4\pi^{2}}.
\end{equation}
Both Eq.~\eqref{eq:pressurepqcd2023} and Eq.~\eqref{eq:pqcdfgpressure} are defined in the massless-quark limit, resulting in three equal quark chemical potentials $\mu_q \equiv \mu_{B}/3$. Three-flavor massless quark matter is charge neutral and in weak equilibrium when $n_u=n_d + n_s$ and $\mu_u=\mu_d=\mu_s$. The electron chemical potential in this limit is zero.
\begin{figure}[t]
    \centering
    \includegraphics[width=\columnwidth]{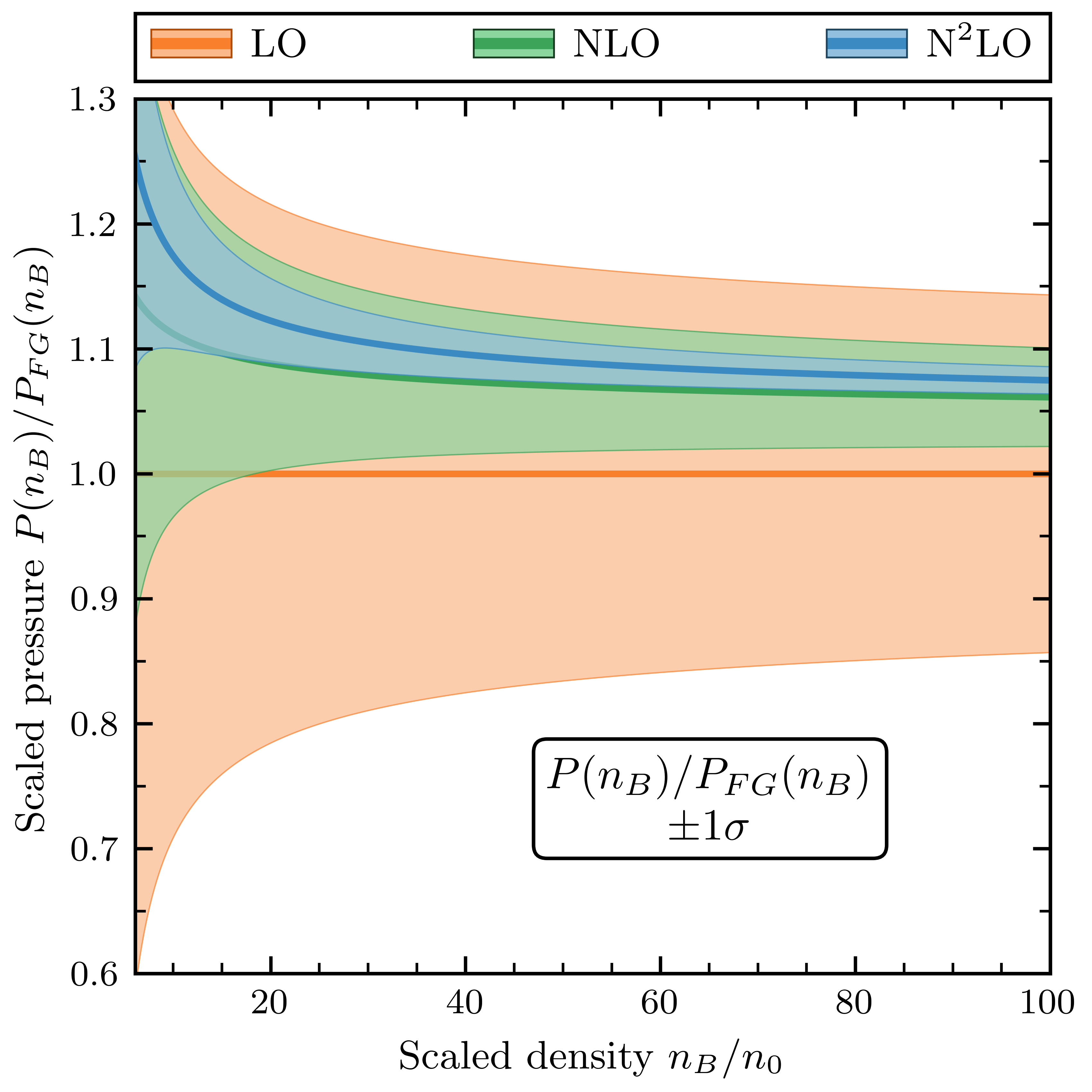}
    \caption{Order-by-order results for the pressure $P(n_B)$ of three-flavor, massless quark matter from pQCD, scaled with the Fermi gas pressure $P_{FG}(n_B)$, plotted against the scaled baryon number density $n_B/n_0$. The 1$\sigma$ uncertainties are obtained using the BUQEYE truncation error model~\cite{Melendez:2019izc}.}
    \label{fig:pqcd_anm}
\end{figure}

 In the limit of asymptotically large chemical potentials, the pQCD EOS calculation using two massless quarks and one massive (strange) quark ($N_f=2+1$) converges to the result with three massless quarks ($N_f=3$). Here, a quark chemical potential of $0.8 \GeV$ corresponds to $n_B = 40n_0$, and at this value of $\mu_q$ the difference in the scaled quark number density $n(\mu_q)/n_{FG}(\mu_q)$ for the $N_f=3$ and $N_f=2 + 1$ cases is negligible. Hence, for $n_B \geqslant 40 n_0$ either calculation can be used to determine the EOS from pQCD. Meanwhile, for $N_f=3$, our second, lower cutoff density of $20n_0$ corresponds to a quark chemical potential of $0.705 \GeV$. The difference between the $N_f=3$ and $N_f=2+1$ values of $n(\mu_q)$ at this $\mu_q$ is $0.02~n_{FG}$~\cite{Kurkela:2009gj}. This difference is half the size of the N$^2$LO $n(\mu)$ truncation error for the three-flavor massless case: $0.04~n_{FG}$. We therefore deem the three-flavor massless quark description of pQCD sufficient for this study at densities $\geqslant 20n_0$.

We require pQCD described in terms of $n_{B}$ when we perform our model mixing procedure. Hence we take Eq.~\eqref{eq:pressurepqcd2023} and convert it to $P(n_B)/P_{FG}(n_B)$ via the Kohn-Luttinger-Ward inversion~\cite{Kohn:1960zz, Luttinger:1960ua} (see Appendix~A of Paper I for details); we obtain
\begin{align}
    \label{eq:scaledpqcdpressuren}
    \frac{P(n_B)}{P_{FG}(n_B)} &= 1 + \frac{2}{3\pi}\alpha_{s}(\Lambdabar_{FG}) + \frac{8}{9\pi^{2}}\alpha_{s}^{2}(\Lambdabar_{FG}) \notag \\ &\quad- \frac{9}{3\pi^{2}}c_{2}(\mu_{FG})\alpha_{s}^{2}(\Lambdabar_{FG}) \nonumber \\
    &\quad- \frac{\beta_{0}}{3\pi^{2}}\alpha_{s}^{2}(\Lambdabar_{FG}),
\end{align}
where $c_{2}$ is defined as
\begin{align}
    c_{2}(\mu_{FG}) &= \frac{1}{3}\bigg[a_{2,1}\ln{\left(\frac{3\alpha_{s}(\Lambdabar_{FG})}{\pi}\right)} \nonumber \\ 
    &\quad+ a_{2,2}\ln{\left(\frac{\Lambdabar_{FG}}{2\mu_{FG}}\right)} + a_{2,3}\bigg].
\end{align}
We take the Fermi gas chemical potential as $\mu_{FG} = (\pi^{2}n_B/3)^{1/3}$, and $\beta_{0} = 11 - 2N_{f}/3 = 9$. In the above expressions, all quantities now depend only on $\mu_{FG}$, and hence are straightforwardly computed at a particular $n_B$.
    
In Paper~I, we used the BUQEYE truncation error model~\cite{Melendez:2019izc} to obtain order-by-order uncertainties for the EOS from pQCD. Here, we use the same prescription, calculating the truncation error bands at each order through N$^2$LO. The results of Eq.~\eqref{eq:scaledpqcdpressuren} and the corresponding 1$\sigma$ uncertainties are plotted in Fig.~\ref{fig:pqcd_anm}.


\section{Bayesian model mixing framework} \label{sec:bmm}

In this work we use GPs to perform model mixing. For full details of our formalism, see Sec.~III~A of Paper~I. Here we provide a short overview, defining the EOSs of $\chi$EFT and pQCD in the form
\begin{equation}
    \label{eq:predictions_curve}
    Y^{(i)}(x) = F(x) + \delta Y^{(i)}(x),
\end{equation}
where $Y^{(i)}$ is the pressure of either $\chi$EFT or pQCD, while $F$ is the pressure in the underlying theory that we wish to infer. Since in what follows we perform model mixing in two different input spaces---density and log density---we keep the input space $x$ generic in this section when discussing the model mixing prior and posterior. 

The uncertainties on the EOS from pQCD are given by the BUQEYE truncation error model. This means that we can write the uncertainties on this EOS as 
\begin{equation}
    \label{eq:GPerrormodel}
    \delta Y^{(\textrm{pQCD})}(n_B) \sim \mathcal{GP}[0, \kappa_{y}^{(\textrm{pQCD})}(n_B,n_B')],
\end{equation}
where $\kappa_{y}^{(\textrm{pQCD})}(n_B,n_B')$ is the covariance function defining the correlations between points $n_B$ and $n_B'$ of the input space. In our case, $y = P(n_B)$. For the EOS from $\chi$EFT, 
the kernel corresponds to the mixed partial derivative of Eq.~\eqref{eq:nsm_cov_energy}, including appropriate density factors (see Eq.~\eqref{eq:pressure_nsm}):
\begin{equation}
    \label{eq:nsm_cov_pressure}
    \kappa_{y}^{(\chi \textrm{EFT})}(n_B,n_B') = n_{B}^{2}n_{B}'^{2}~\partial_{n_B}\partial_{n_B'}\kappa_{E_{\textrm{NSM}}}(n_B,n_B').
\end{equation}
This kernel is evaluated via numerical differentiation and then provides the GP representation of the uncertainties for the $\chi$EFT pressure as a function of baryon density:
\begin{equation}
    \label{eq:deltaychiraleft}
    \delta Y^{(\chi \textrm{EFT})}(n_B) \sim \mathcal{GP}[0, \kappa_{y}^{(\chi \textrm{EFT})}(n_B,n_B')].
\end{equation}

We also require a prior on the underlying theory; we use a GP representation there too,
\begin{equation}
    \label{eq:theorygpprior}
    F(n_B) \given I \sim \mathcal{GP}[m(x), \kappa_{f}(x,x')],
\end{equation}
where the mean $m(x)$ and covariance functions $\kappa_{f}(x,x')$ define the prior and $I$ is the information we employ in choosing them. In the following sections we exemplify various choices that can be made for these functions and show how they affect the results of the model mixing inference. 

We can then obtain a GP representation of the underlying theory for $F$~\cite{rasmussen2006gaussian, Semposki:2024vnp}, incorporating the prior information~\eqref{eq:theorygpprior}, and the pQCD and $\chi$EFT training data, by writing the joint Gaussian distribution of $F$ at a set of evaluation points in the $x$ input space. We denote this set of Gaussian random variables by $F_e$. The resulting distribution of $F_e$, given a training set of $N$ model evaluations $\{(y_t)_i\}_{i=1}^N$, the corresponding covariance matrix $K_{y,tt}$, and the prior on $F$, is
\begin{align}
    F_e \given \{(y_t)_i\}_{i=1}^N, K_{y,tt}, I \sim \mathcal{N}[\mu_e, \Sigma_{ee}], 
\end{align}
 with $\mu_e$ and $\Sigma_{ee}$ given by~\cite{rasmussen2006gaussian}
\begin{align}
    \label{eq:evaluationmean}
    \mu_e & = K_{f,et} (K_{f,tt} + K_{y,tt})^{-1} (y_t - m_t) + m_t, \\
    \label{eq:evaluationcov}
    \Sigma_{ee} & = K_{f,ee} - K_{f,et} (K_{f,tt} + K_{y,tt})^{-1} K_{f,te}.
\end{align}
Here the mean function $m_t$ is evaluated at the training data. Meanwhile, the subscripts $y$ and $f$ of the covariance matrices $K$ indicate the data or prior covariance matrix, while the subscripts on $K$ after the comma denote the set of (training or evaluation) points it is constructed at.

This BMM framework computes the scaled pressure, $P(n_B)/P_{FG}(n_B)$. Once we have this in hand, we can calculate the speed of sound squared via
\begin{equation}
    \label{eq:cs2}
    c_{s}^{2}(n_B) = \frac{1}{\mu_B} \frac{\partial P}{\partial n_B},
\end{equation}
where $\mu_B$ is the baryon chemical potential.
A detailed discussion of our workflow to obtain $\mu_B$, and hence the speed of sound, is presented in Sec.~III~B of Paper~I.


\section{Stationary Gaussian Processes} \label{sec:stationaryGPs}

In Paper~I, we used the squared-exponential RBF kernel to guide interpolation of the EOS in the density range between $\chi$EFT and pQCD. In Sec.~\ref{sec:stationarykernels}, we introduce three other stationary kernel classes to describe our model mixing prior, and explain the choices we make for their hyperparameter priors. In Sec.~\ref{sec:stationaryresults}, we present the results for the scaled pressure and speed of sound when the squared-exponential RBF kernel is used. We discuss the resulting mass-radius relation for a neutron star in Sec.~\ref{sec:mrstationaryresults}. Finally, in Sec.~\ref{sec:otherkernels}, we investigate the Mat\'ern and rational quadratic kernels and present the corresponding results for the scaled pressure. 


\subsection{Kernels and hyperpriors} \label{sec:stationarykernels}

In this section we consider three classes of stationary correlation functions: the squared-exponential, the Mat\'ern, and the rational quadratic. They are defined in Ref.~\cite{rasmussen2006gaussian} and below. In each case we adopt a zero mean function for our initial analysis. We employ the Python package \texttt{scikit-learn} to use these kernels in our BMM framework~\cite{scikit-doc, scikit-learn}. Each of them are multiplied by a constant kernel to establish a marginal variance, $\bar{c}^{2}$, in the covariance structure, yielding
\begin{equation}
    \label{eq:covariancefunctiongeneral}
    \kappa(x,x'; \bm{\theta}) = \bar{c}^{2} k(x,x';\bm{\theta}),
\end{equation}
where all hyperparameters of the covariance functions are contained in $\bm{\theta}$. While we define our individual models and their uncertainties~\eqref{eq:GPerrormodel} and~\eqref{eq:deltaychiraleft} in $n_B$ space, we perform all of our model mixing in this section in 
$\ln(n_B/1\,{\rm fm}^{-3})$  
space. In this section this, not $n_B$, is the input space considered in Eq.~\eqref{eq:theorygpprior}; this makes the model mixing more stable across the large intermediate region between the two theories.

Generally, covariance functions of GPs contain one or more hyperparameters that are learned from the training data in a given problem. Before defining the kernel structures of each stationary case explored, we establish the general prior forms that we use on the hyperparameters of each kernel. We take a truncated normal distribution for the marginal variance, represented by
\begin{align}
    \label{eq:marginalvariancegeneral}
    \bar{c}^2 \given I &\sim \mathcal{U}(a=0.15, b=6.25) \nonumber \\ 
    &\quad\times \mathcal{N}\left(\mu=1.0, \sigma^{2}=0.25^{2}\right).
\end{align}
Equation~\eqref{eq:marginalvariancegeneral} constitutes a rather broad prior for $\bar{c}^2$, and centers it around naturalness. This hyperprior choice remains the same regardless of the kernel structure; we have no physics reason for changing our priors on this quantity as we change the correlation function $k$.

Each of the correlation functions investigated in this section contains a length scale, $\ell$. This parameter generally encodes the distance in the input space over which correlations in the data exist. For the current analysis, we use a truncated normal distribution as our prior on the length scale:
\begin{equation}
    \label{eq:hyperpriorgeneral}
    \ell \given I \sim \mathcal{U}(a_\ell, b_\ell) \times \mathcal{N}\left(\mu_\ell, \sigma_\ell^{2}\right),
\end{equation}
where $a_{\ell},b_{\ell}$ are the hard cutoffs of the uniform distribution, and $\mu_{\ell}, \sigma^{2}_{\ell}$ are the mean and variance of the normal distribution, respectively. The parameters chosen for the hyperpriors on each kernel's length scale are given in Table~~\ref{tab:stationary_priors}.

We begin our covariance function analysis with the squared-exponential RBF (hereafter just ``RBF'') kernel, which has the structure
\begin{equation}
    \label{eq:rbfkernel}
    k_{\textrm{RBF}}(x,x'; \ell) = \exp\left( \frac{-(x-x')^{2}}{2\ell^{2}} \right).
\end{equation}
Its only hyperparameter is the length scale $\ell$, which, for the case of this kernel, directly describes the correlation length of the function. We place a hyperprior on $\ell$ using Eq.~\eqref{eq:hyperpriorgeneral}. We select the range $[a_{\ell},b_{\ell}]$ by recalling that neither $\chi$EFT nor pQCD training data should, \textit{a priori}, influence predictions in the other theory's region, since we initially assume no inter-model correlations between the two sets of training data used. This then implies that the correlation falloff in Eq.~\eqref{eq:rbfkernel} must be reasonably rapid, so that there is minimal influence from a particular theory by the time the GP reaches the opposite edge of the intermediate region. We therefore choose the range of $[a_{\ell},b_{\ell}]$ to correspond to an initial expectation of $3\%-5\%$ correlation between \ChiEFT~at $n_B=2n_0$ and pQCD at densities of $2n_0$ and $20n_0$ or $40n_0$, depending on the choice of pQCD lower limit. The mean value of the truncated normal is taken to be around the lower bound of this interval, to favor smaller correlations. 

\begin{figure*}[t]
        \centering
        \includegraphics[width=\textwidth]{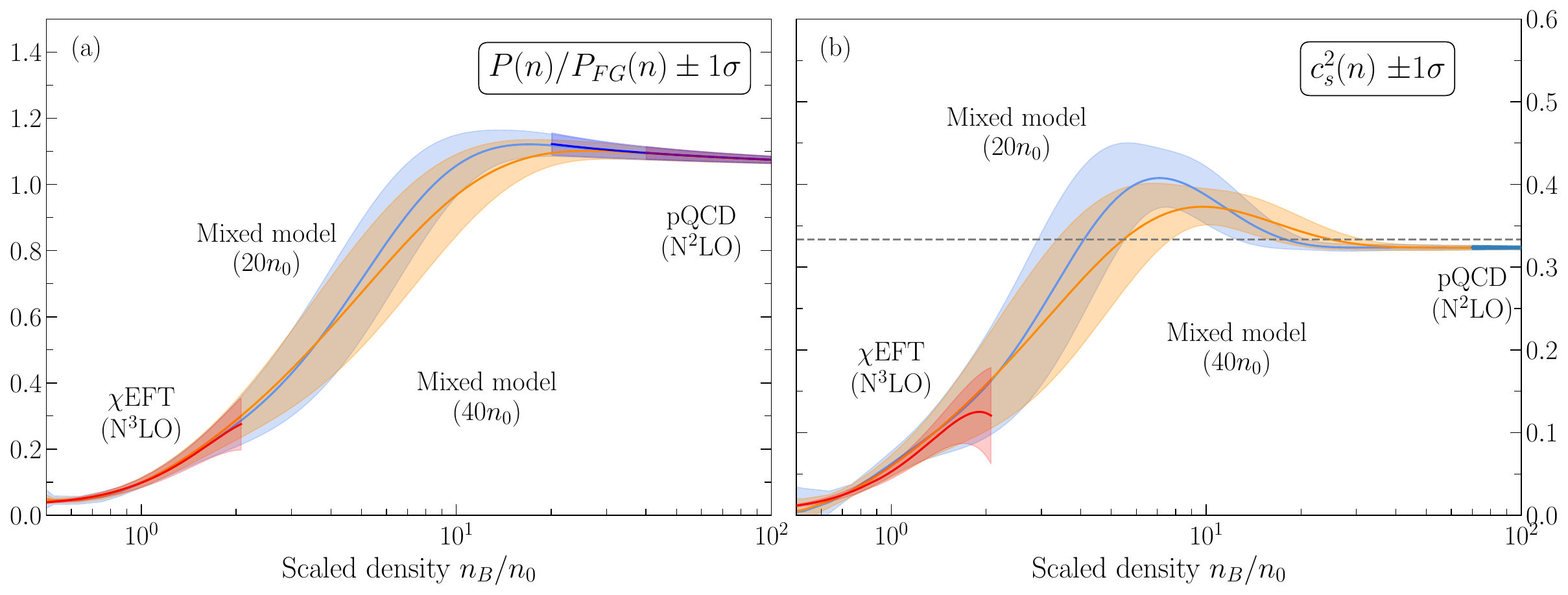}
        \caption{(a) The 68\% bands of the mixed-model EOS for the scaled pressure $P(n_B)/P_{FG}(n_B)$, computed with both pQCD lower limits: $20n_0$ (blue) and $40n_0$ (orange). The 68\% intervals of the input theories---$\chi$EFT (red band) and pQCD at both cutoffs $20n_0$ and $40n_0$ (dark blue and purple bands, respectively)---are also shown. (b) The speed of sound squared for the mixed-model EOS at both pQCD lower limits as in (a). We note that the speed of sound squared approaches the conformal limit, $c_{s}^{2}(n_B) \rightarrow 1/3$ from below, in the region where pQCD dominates ($\geqslant 20n_0-40n_0$), and also contains a gradual peak in both cases.}
        \label{fig:pressure_cs2_stationary}
    \end{figure*}

\begin{table}[t]
\caption{The choices for the hyperpriors on the length scales of each stationary covariance function in Sec.~\ref{sec:stationarykernels} for both pQCD lower limits, $20n_0$ and $40n_0$. The form of the prior used for each is given in Eq.~\eqref{eq:hyperpriorgeneral}. $\mu_\ell$ ($\sigma_l$) corresponds to the mean (standard deviation) of the normal distribution over $\ell$. The bounds in the last column are the values $[a_{\ell}, b_{\ell}]$. These quantities are expressed in $\ln{(n_B/1 \textrm{fm}^{-3})}$ space.}
\centering
\begin{tabular}{c|c|c|c|c}
   pQCD & \multirow{2}{*}{Kernel} & \multicolumn{3}{c}{Length scale, $\ell$} \\
   \cline{3-5}
   lower limit & & $\mu_\ell$ & $\sigma_\ell$ & Bounds \\
   \hline
   \multirow{4}{*}{$20n_0$} & RBF & 0.89 & 0.15 & [0.87, 0.94] \\
    & Mat\'ern 3/2 & 0.76 & 0.15 & [0.75, 0.84] \\
    & Mat\'ern 5/2 & 0.8 & 0.15 & [0.78, 0.87] \\
    & RQ ($\alpha=2$) & 0.55 & 0.15 & [0.53, 0.62] \\
   \hline
   \multirow{4}{*}{$40n_0$} & RBF & 1.15 & 0.15 & [1.13, 1.22] \\
    & Mat\'ern 3/2 & 0.99 & 0.15 & [0.97, 1.1] \\
    & Mat\'ern 5/2 & 1.03 & 0.15 & [1.01, 1.1] \\
    & RQ ($\alpha=2$) & 0.7 & 0.15 & [0.69, 0.8] \\
   \hline
\end{tabular}
\label{tab:stationary_priors}
\end{table}

We follow a very similar procedure for the Mat\'ern kernel class, implementing both the $\nu=3/2$ and $\nu=5/2$ functions. GPs with the Mat\'ern kernel are not infinitely differentiable, in contrast to those with the RBF kernel. This kernel is defined by~\cite{duvenaud_PhD_2014}
\begin{equation}
    k_{\textrm{Mat\'ern}}(x,x'; \nu, \ell) = \frac{2^{1-\nu}}{\Gamma(\nu)} \left(\frac{\sqrt{2\nu}r}{\ell}\right)^{\nu} K_{\nu}\left(\frac{\sqrt{2\nu}r}{\ell}\right).
\end{equation}
Here we also have a hyperparameter called the length scale $\ell$; however, it does not directly map to the length scale of the squared-exponential RBF kernel, because of the different covariance structures of the two kernels~\cite{rasmussen2006gaussian}. 

We also implement the rational quadratic (RQ) kernel, which has the form
 \begin{equation}
    k_{\textrm{RQ}}(x,x'; \alpha, \ell) = \left( 1 + \frac{r^{2}}{2\alpha\ell^{2}} \right)^{-\alpha}.
\end{equation}
$k_{\textrm{RQ}}$ has two hyperparameters, and can be interpreted as a mixture of squared-exponential RBF functions with different length scales~\cite{rasmussen2006gaussian}. The scale-mixture parameter $\alpha$ controls the similarity of the rational quadratic to the RBF kernel; in the limit $\alpha \rightarrow \infty$, $k_{\textrm{RQ}} \rightarrow k_{\textrm{RBF}}$. Here we set $\alpha=2$. This leaves $\ell$, the length scale, as the only free hyperparameter, as with the other two kernel classes. We assign a prior using Eq.~\eqref{eq:hyperpriorgeneral} as before, noting again that $\ell_{\textrm{RQ}}$ also does not map directly to $\ell_{\textrm{RBF}}$.

Optimizing the hyperparameter values for the length scales of the kernel classes chosen often yields a value at the edge of the prior range. If there are no intermediate data in the region $2n_0 \leqslant n_B \leqslant 40 n_0$---or, more generally, if any data being used for EOS inference in that region are ``$m$-constant'', i.e., differ from the mean function by a constant---then the exponential part of the likelihood has no discriminating power as regards the kernel length scale and the optimal kernel length scale found will be as large as possible~\cite{JMLR:v24:22-1153}. This happens because the normalized likelihood includes a pre-factor of $1/\sqrt{{\rm det} k}$ that functions as a ``model complexity" penalty since it favors kernels with a small determinant~\cite{rasmussen2006gaussian}. In our case, if no hyperpriors are implemented, this results in the pQCD EOS dominating the correlation structure in the GP kernels over as much of the input space as possible. This outcome is unphysical and we introduce the aforementioned hyperpriors on the kernel hyperparameters to prevent it. The MAP (\textit{maximum a posteriori}) values that result for the $\ell$ and $\cbar^2$ of each kernel with the hyperpriors in place are listed in Table~\ref{tab:stationary_priors_MAP}. Owing to the tendency of gradient-descent optimizers to get stuck in local extrema, we restart our optimizer in $\approx 10,000$ different random locations to ensure that the solution for the hyperparameters has converged. Repeating this procedure several times produces variation in the length scale of $\pm 0.01$ and in the marginal standard deviation of at most $\pm 0.1$. These variations in the gradient-descent solution do not constitute significant changes in the mixed-model EOS.


\subsection{EOS results for the squared-exponential RBF kernel} \label{sec:stationaryresults}

\begin{table}
\caption{The MAP values of the length scales and marginal variances of each stationary covariance function in Sec.~\ref{sec:stationarykernels} for both pQCD lower limits, $20n_0$ and $40n_0$, using the zero mean function. $\ell_{\textrm{MAP}}$ corresponds to the MAP value of the length scale, and $\bar{c}^{2}_\textrm{MAP}$ corresponds to the MAP value of the marginal variance. Note that the length scale is expressed in $\ln{(n_B/1 \textrm{fm}^{-3})}$ space.}
\centering
\begin{tabular}{c|c|c|c}
   pQCD & \multirow{2}{*}{Kernel} & \multirow{2}{*}{$\ell_{\textrm{MAP}}$} & \multirow{2}{*}{$\bar{c}^{2}_{\textrm{MAP}}$} \\
   lower limit & & & \\
   \hline
   \multirow{4}{*}{$20n_0$} & RBF & 0.94 & 0.72 \\
    & Mat\'ern 3/2 & 0.83 & 0.15 \\
    & Mat\'ern 5/2 & 0.86 & 0.15 \\
    & RQ ($\alpha=2$) & 0.62 & 0.23 \\
   \hline
   \multirow{4}{*}{$40n_0$} & RBF & 1.2 & 0.79 \\
    & Mat\'ern 3/2 & 1.1 & 0.15 \\
    & Mat\'ern 5/2 & 1.1 & 0.15 \\
    & RQ ($\alpha=2$) & 0.80 & 0.25 \\
   \hline
\end{tabular}
\label{tab:stationary_priors_MAP}
\end{table}

To construct the EOS for NSM using the EOSs from $\chi$EFT and pQCD, we first define a data set, with full covariances, that we use to train our GP prior. We extract 4 points from the scaled pressure of $\chi$EFT, and 7 (5) points from the scaled pressure of pQCD using a lower limit of $20n_0$ ($40n_0$). Employing additional training points produces an ill-conditioned inverse in Eq.~\eqref{eq:evaluationcov}. 

Our results using the stationary, squared-exponential RBF kernel to model mix $\chi$EFT and pQCD for charge-neutral, $\beta$-equilibrated matter are presented in Fig.~\ref{fig:pressure_cs2_stationary}. There we show the mixed-model EOS in terms of the scaled pressure $P(n_B)/P_{FG}(n_B)$, and the consequent speed of sound squared, $c_{s}^{2}(n_B)$ from~\eqref{eq:cs2}. 
The stationary GP solution found has a small correlation between the $P(n_B)$'s at densities in the $\chi$EFT and pQCD regions. It thus mostly reproduces the input $\chi$EFT and pQCD error bands. We do, however, see a loss of precision for $n_B = 0.5n_{0} - n_{0}$ in both the pressure and the speed of sound. At high densities
the speed of sound approaches the conformal limit from below, as expected in the region dominated by pQCD~\cite{Semposki:2024vnp}, and we see a stable ($c_{s}^{2} \geqslant 0$) and causal ($c_{s}^{2} \leqslant 1$) EOS is produced.


\subsection{\texorpdfstring{$M-R$}{M-R} constraints for the squared-exponential RBF kernel} \label{sec:mrstationaryresults}

To study the implications of our $\beta$-equilibrated EOSs for the structure of static, spherically symmetric neutron stars, we solve the Tolman-Oppenheimer-Volkoff (TOV) equations~\cite{tolman1939, oppenheimer1939}
\begin{subequations} \label{eq:toveq}
\begin{align}
    \dv{p}{r} &= -G(\varepsilon + p) \frac{M(r) + 4\pi r^3 p}{r(r-2GM(r))}, \label{eq:tovpres} \\
    \dv{M}{r} &= 4\pi r^2 \varepsilon, \label{eq:tovmass}
\end{align}
\end{subequations}
where $G$ is the gravitational constant and $\varepsilon$ the energy density of the system. The relationship between $\varepsilon$ and $p$ is the mixed-model EOS. The TOV equations are solved in scaled form using standard Runge-Kutta methods~\cite{Piekarewicz2017, Lalit:2024vmu} over a range of central pressures (or equivalently, central densities). The central pressure acts as an initial condition and we then solve $p(r)$ until we reach the radius at which $p=0$. That radius, and the enclosed mass determine one point on the neutron star mass-radius relation curve.

Since we need a crust EOS to perform this calculation, we simply adopt the low-density EOS from Refs.~\cite{Negele:1972zp, Baym:1971pw} and match this with $\chi$EFT at $n_B = 0.08 \fmiq$. This is done by concatenating the two EOSs at this density, i.e., masking the pressures and energy densities from our mixed-model EOS in favor of the crust EOS below $0.08 \fmiq \approx 0.5n_0$. Since this crust EOS model lacks quantified uncertainties, we attach the same crust EOS curve to each draw from our mixed-model GP for $P(n_B)/P_{FG}(n_B)$.

\begin{figure}[t]
    \centering
    \includegraphics[width=\columnwidth]{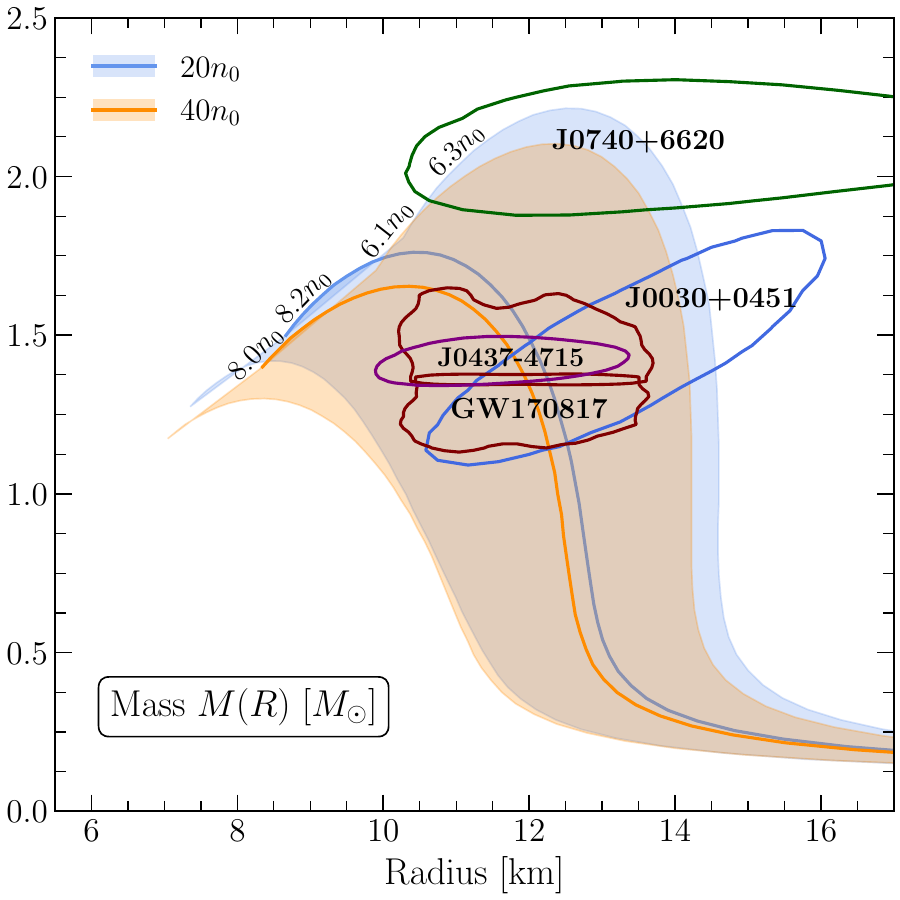}
    \caption{The mass-radius relation of a static, spherically symmetric neutron star for both pQCD lower limits: $20n_0$ (blue) and $40n_0$ (orange). Each shaded region corresponds to the 68\% credibility interval of the EOS translated into $M-R$ plane. We calculate the central densities of each mean and upper 68\% interval and label each curve with the corresponding values. We also overlay the contours from the LIGO-Virgo (GW170817) and NICER (J0437-4715, J0740+6620, and J0030+0451) scientific collaborations at the 90\% credibility level~\cite{choudhury_2024_13766753, Choudhury:2024xbk, Riley:2019yda, Riley:2021pdl, miller_2019_3473466, miller_2021_4670689, Miller:2021qha, Miller:2019cac, LIGOScientific:2018cki}.}
    \label{fig:mr_comparison_stationary}
\end{figure}

Figure~\ref{fig:mr_comparison_stationary} displays our 68\% credibility intervals for the $M-R$ relation for two different mixed-model EOSs obtained with the squared-exponential RBF kernel: one has a pQCD lower limit of $20n_{0}$ (blue) and one a lower limit of $40n_{0}$ (orange). Overlaid on these 68\% regions are the 
current constraints from LIGO-Virgo and NICER data at the 90\% credibility level~\cite{choudhury_2024_13766753, Choudhury:2024xbk, Riley:2019yda, Riley:2021pdl, miller_2019_3473466, miller_2021_4670689, Miller:2021qha, Miller:2019cac, LIGOScientific:2018cki}.
Our 68\% intervals encompass significant amounts of the 90\% credible regions corresponding to NICER and LIGO-Virgo observations. We therefore conclude that our mixed model is consistent with the NICER and LIGO-Virgo measurements plotted in Fig.~\ref{fig:mr_comparison_stationary}. For the two mean curves and the curves at the upper edge of the two 68\% envelopes we indicate the central density corresponding to the maximum mass on the curve. The curves corresponding to the mean and the upper edge of the 68\% region give realistic central densities for neutron stars, in the range $\approx 6n_0-8n_0$~\cite{Chatziioannou:2024tjq, Koehn:2024set}. Both 68\% regions have some overlap with the region  $M > 2M_{\odot}$ and are thus consistent with observations, such as those of Ref.~\cite{Romani:2022jhd}. We do not achieve any significant constraints on the neutron star radius from this kernel; Fig.~\ref{fig:mr_comparison_stationary} shows a range of radii $\sim 10-13$ km from the 68\% bands on the EOS propagated to $M-R$ space. 


\subsection{Results for other stationary kernels}\label{sec:otherkernels}

\begin{figure}
        \centering
        \includegraphics[width=\columnwidth]{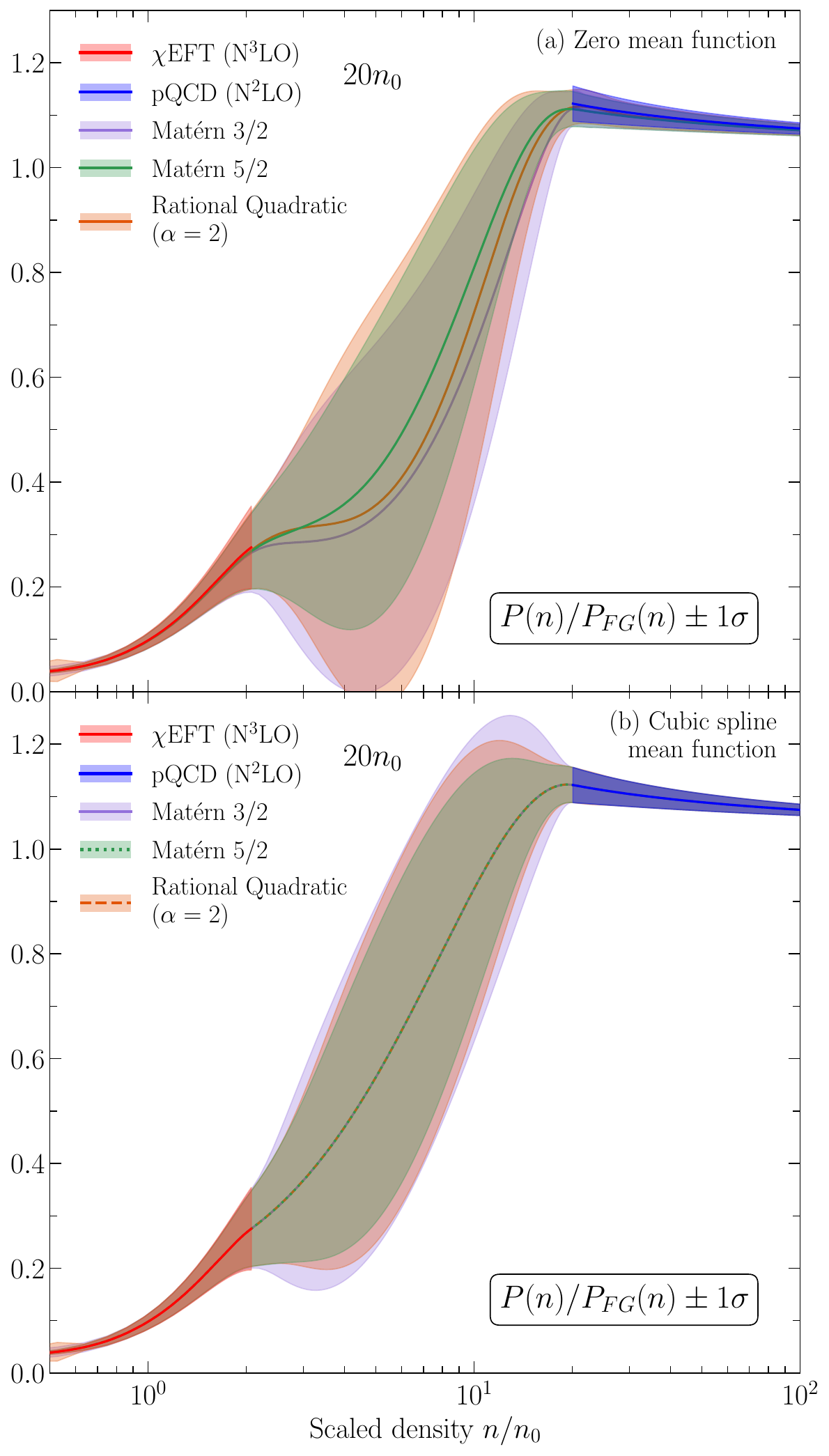}
        \caption{(a) The mixed-model EOS from various stationary kernels: the Mat\'ern 3/2 (purple), Mat\'ern 5/2 (green), and rational quadratic with a choice of $\alpha=2$ (orange). Here we use the pQCD lower limit at $20n_0$, and take a zero mean function for our GP prior when mixing the models. (b) Same as (a) but using a cubic spline mean function to interpolate across the intermediate region. Hence, the mean values of each kernel are equal in the region between $\chi$EFT and pQCD, as is shown by the dotted (Mat\'ern 5/2), dashed (RQ), and solid (Mat\'ern 3/2) mean curves lying exactly on top of one another.}
        \label{fig:stationary_kernel_comparison}
    \end{figure}

\begin{figure*}[t]
    \centering
    \includegraphics[width=\textwidth]{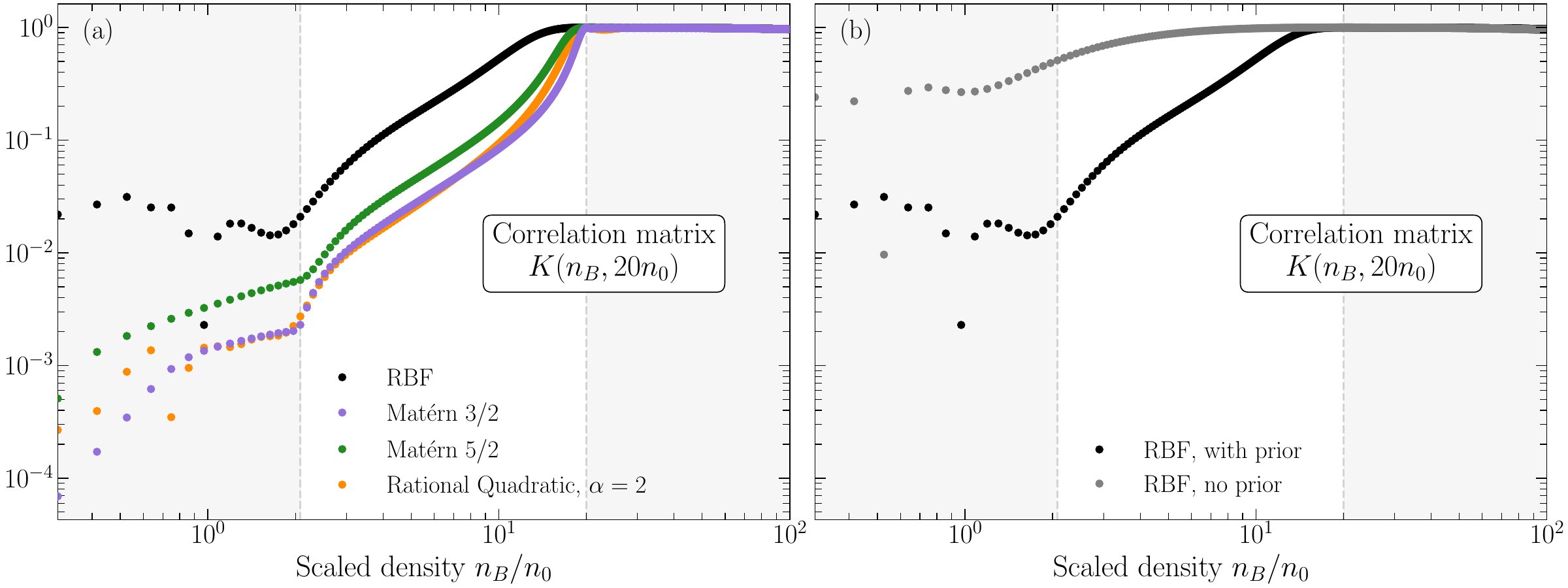}
    \caption{(a) The absolute value of the kernel correlation matrix element $K(n_B, 20n_0)$ as a function of the scaled density $n_B/n_0$ for each of the kernel classes considered, with a zero mean function used. The regions where data is used to train each kernel are shaded in light gray. Here we have implemented the truncated normal prior on the length scale hyperparameter of each kernel, reducing the persistence of correlations from pQCD into the density range of $\chi$EFT. (b) Here we compare the difference between the correlations with the truncated normal prior on the length scale of the squared-exponential RBF kernel (black) and the case without this prior implemented (gray). We note the large difference between the correlations in the two scenarios below $n_B = 20 n_0$.}
    \label{fig:correlations}
\end{figure*}

To assess the sensitivity to the choice of GP kernel, we now employ the Mat\'ern and RQ kernels to model mix $\chi$EFT and pQCD. Here we begin again with the zero mean function and train our kernels on the means and covariances from the two theories. For the Mat\'ern kernels, we use data sets corresponding to 30 points from $\chi$EFT and 49 (37) points from pQCD with a lower limit of $20 n_0$ ($40n_0$). For the RQ kernel, we cannot use very many data points due to the GP matrix inversion in Eq.~\eqref{eq:evaluationcov} becoming ill-conditioned, so we choose 7 points from $\chi$EFT and 10 (8) points from pQCD with a lower limit of $20 n_0$ ($40n_0$).

In Fig.~\ref{fig:stationary_kernel_comparison}, we compare the results of the model mixing procedure for the different kernel classes other than the squared-exponential RBF using the pQCD lower limit of $20n_0$. (Very similar behavior occurs for the case of $40n_0$.) We see in Fig.~\ref{fig:stationary_kernel_comparison} that the Mat\'ern 3/2 kernel produces the largest uncertainties, but the uncertainty band is of a comparable size for all the kernels shown there. However, the Mat\'ern 3/2, Mat\'ern 5/2, and Rational Quadratic kernel each produce a much larger uncertainty band in the intermediate region than we saw for the case of the RBF kernel (see Fig.~\ref{fig:pressure_cs2_stationary}). This makes sense in light of the generally less smooth properties of these kernel correlation functions in comparison to the RBF kernel. 

We also see in Fig.~\ref{fig:stationary_kernel_comparison}a that the 68\% uncertainty bands for the Mat\'ern 3/2 and the RQ kernel dip below zero. The derivative $\partial P(n_B)/\partial n_B$ is negative there, so $c_{s}^{2} = \mu^{-1}\partial P(n_B)/\partial n_B < 0$ and the EOS is unstable. To remedy this, we choose to employ a cubic spline mean function instead of the original choice of a zero mean function; choosing this non-zero form allows the GP to ensure that the unscaled pressure is monotonically increasing as a function of density. Implementing this generally positive slope in the scaled pressure certainly ensures this characteristic in the unscaled pressure. We show the resulting EOSs from using the cubic spline mean function in Fig.~\ref{fig:stationary_kernel_comparison}b. Employing this mean function stabilizes EOSs within the entire 68\% interval, and also yields a stiffer EOS overall.

The Mat\'ern 3/2 and RQ kernels produce uncertainties for the mixed-model EOS below $n_0$ that are markedly larger than the input uncertainties from $\chi$EFT. Since the region $n_B \leqslant n_0$ is important in constraining the radius of a neutron star~\cite{Lattimer:2000nx}, we regard the recovery of $\chi$EFT as an important criterion on the usefulness of a particular kernel for this problem. 
The Mat\'ern 5/2 kernel is successful in producing a stable EOS which also is able to describe matter around $0.5n_0-n_0$ reasonably well. The $M-R$ result corresponding to this kernel, however, is less constraining than that of the squared-exponential RBF kernel, owing to the larger uncertainty bands in the intermediate region. 

It might seem surprising, given the different smoothness properties of the four stationary covariance functions tested here, that each produces a similar mixed model. These similarities are due to the comparable correlation coefficients of each kernel, which we examine after the GP updating in Eq.~\eqref{eq:evaluationcov} has occurred, between evaluation points at densities of $2n_0$ and $20n_0$. Figure~\ref{fig:correlations}a compares the resulting correlation matrices for the optimal kernel in each class after the hyperpriors are imposed. 
For all cases except the RBF kernel the correlation coefficients between $n_B=2 n_0$ and $20 n_0$ are $0.2\%-0.6\%$, which is an order of magnitude lower than we originally expected to achieve with our hyperpriors. For comparison, the RBF kernel exhibits a correlation coefficient of $2.0\%$. The GP updating results in significantly lower overall correlations in the conditioned kernel as compared to the prior kernel. The RBF kernel still exhibits the most persistence of correlations between $2n_0$ and $20n_0$, but they are now small enough that the influence of the pQCD data and any concomitant shrinking of $\chi$EFT uncertainty bands is not noticeable.

As previously mentioned, if we do not impose hyperpriors then the very small error bars in the training data from pQCD, together with the correlations that persist down to densities where $\chi$EFT should be the favored model, result in an unphysical decrease in the uncertainties of the mixed model. 
We illustrate the marked effect of the hyperprior on the correlations between $\chi$EFT and pQCD in Fig.~\ref{fig:correlations}b. Figure~\ref{fig:correlations}b compares the correlations for the RBF kernel with and without a hyperprior imposed. Here, the persistence of the correlations induced by pQCD and its tight error bars is evident; the correlations between densities of $2n_0$ and $20n_0$ are an order of magnitude larger when no prior on the length scale is used than when our truncated normal prior \eqref{eq:hyperpriorgeneral} is implemented. While we have only shown this reduction in correlation upon imposition of the hyperprior for the case of the RBF kernel, the hyperprior produces an even more severe reduction for the other kernels considered. 


\section{Non-stationary Gaussian processes: changepoint kernels} \label{sec:changepointGPs}

It is bold to assume that the EOS of strongly interacting matter is described by a GP kernel with a single length scale, i.e., that the variation of $P(n_B)$ has a characteristic scale in $\ln{(n_B)}$ that is consistent all the way from pQCD matter through to nuclear matter. In this section, we show how to account for the possibility that strongly interacting matter is instead described by a density-dependent, changepoint kernel~\cite{Garnett:2010rom}, built from two individual (stationary) kernels for the EOSs from $\chi$EFT and pQCD, and joined via a chosen mixing function. We describe the prior and mixing function choices, as well as the formalism of the kernel, in the following section. We then discuss the inclusion of exogenous data in the intermediate density region in Sec.~\ref{sec:mockdata}, and show results for the EOS and the corresponding $M-R$ curves in Sec.~\ref{sec:changepointresults}.


\subsection{Formalism and prior choices} \label{sec:changepointformalism}

When employing a changepoint kernel in our BMM framework, the form of the GP for $F$ is still described by Eq.~\eqref{eq:theorygpprior}; the difference appears purely in the covariance function, $\kappa_f$, which we now write generally as
\begin{align}
    \label{eq:changepointkernel}
    \kappa_{f}(x,x'; \ell_{1}, \ell_{2}, \bar{c}_{1}^{2}, \bar{c}_{2}^{2}, \bm{\xi}) &= [1 - \alpha(x, \bm{\xi})] 
    \nonumber \\
    &\quad\times \kappa_{1}(x,x'; \bar{c}_{1}^{2}, \ell_{1}) [1 - \alpha(x', \bm{\xi})] \nonumber \\
    &\quad+ \alpha(x, \bm{\xi}) \kappa_{2}(x, x'; \bar{c}_{2}^{2}, \ell_{2}) \alpha(x', \bm{\xi}).
\end{align}
Here we define $\ell_{i}$ and $\bar{c}_{i}^{2}$ as the length scales and marginal variances of kernels $\kappa_i$, where $i = 1,2$, and $\bm{\xi}$ is the vector of hyperparameters of the chosen mixing function, $\alpha(x;\bm{\xi})$. The $x$ dependence of this mixing function highlights the non-stationarity of this kernel. $\kappa_1$ and $\kappa_2$ are kernels that are assumed to be known; $\kappa_i$ are the covariance functions of each model in region $i$. Hence, the hyperparameters of these kernels are fixed. The parameter vector $\bm{\xi}$ then denotes the varying hyperparameters of the mixing function that will be determined during the GP training procedure.

In our case the input space is $x=n_B$, while $\kappa_1$ and $\kappa_2$ are the squared-exponential RBF kernels for $\chi$EFT and pQCD, respectively, with the length scales and marginal variances determined previously. We fit these kernels in $n_B$ space, instead of in the $\ln{(n_B)}$ space that was used throughout Sec.~\ref{sec:stationaryGPs} and Paper~I, as we find the changepoint kernel eliminates the need for the latter. The differences between the GP fit and the original covariance matrix of Refs.~\cite{Drischler:2020hwi,Drischler:2020yad} are $< 1$\% throughout this density range. Hence, the RBF kernel provides a reasonable approximation for the $\chi$EFT and pQCD probability density functions (pdfs) in their respective regions.

We invoke a smooth transition between $\chi$EFT and pQCD. To achieve this through the changepoint kernel~\cite{duvenaud_PhD_2014}, we select a sigmoid mixing function, given by
\begin{align}
    \label{eq:sigmoid}
    \alpha_{\textrm{sigmoid}}(x; x_{0}, w) &= \left[ 1 + \exp(-\frac{(x-x_{0})}{w}) \right]^{-1}.
\end{align}
This function is parametrized by $x_{0}$ and $w$, which correspond to the \textit{changepoint} and the \textit{width}, respectively. These hyperparameters indicate, respectively, the point at which the weight of $\kappa_1$ and $\kappa_2$ in the model mixing are exactly equal, and how the slope of the change in these weights is modeled as one moves away from this halfway point.\footnote{We note that the hyperbolic tangent mixing function is another reasonable choice, but since this form is the same as the sigmoid (with an appropriately rescaled width), we only consider the sigmoid function for the remainder of this work.}

\begin{table}[t]
\caption{Choices for the hyperpriors on the changepoint and width hyperparameters in the changepoint kernel using a sigmoid mixing function, for both pQCD lower limits ($20n_0$ and $40n_0$). The form of the prior for each case is that of Eq.~\eqref{eq:hyperpriorchangepoint}; the $\mu$ and $\sigma$ here are the mean and standard deviation of the normal distribution, while the bounds in the last column correspond to $[a,b]$ in Eq.~\eqref{eq:hyperpriorchangepoint}.}
    \centering
    \begin{tabular}{c|c|c|c}
        \hline
       \multicolumn{4}{c}{Changepoint $x_{0}$ [fm$^{-3}$]} \\
       \hline
       pQCD & \multirow{2}{*}{Mean $\mu_{x_0}$} & \multirow{2}{*}{Std.~dev. $\sigma_{x_0}$} & \multirow{2}{*}{Bounds} \\
       lower limit & & & \\
       \hline
       $20n_{0}$ & 0.98 & 0.33 & [0.49, 3.1] \\
       $40n_{0}$ & 0.98 & 0.33 & [0.49, 6.2] \\
       \hline
       \multicolumn{4}{c}{Width $w$ [fm$^{-3}$]} \\
        \hline
       pQCD & \multirow{2}{*}{Mean $\mu_w$} & \multirow{2}{*}{Std.~dev. $\sigma_w$} & \multirow{2}{*}{Bounds} \\
       lower limit & & & \\ 
       \hline
       $20n_0$ & 0.16 & 0.16 & [0.05, 0.32] \\
       $40n_0$ & 0.16 & 0.16 & [0.05, 0.32] \\
       \hline
    \end{tabular}
    \label{tab:prior_choices}
\end{table}
    
We then place appropriate priors on these two hyperparameters to constrain them to physically acceptable ranges. For this we again use the truncated normal distribution
\begin{equation}
    \label{eq:hyperpriorchangepoint}
    \xi_{i} \given I \sim \mathcal{U}(a_i, b_i) \times \mathcal{N}(\mu_i, \sigma_i^{2}),
\end{equation}
We pick suitable ranges for the uniform priors that reflect the need to keep the changepoint between the two model regions: $2n_0 \leqslant x_{0} \leqslant 40 n_0$ (or $20 n_0$ if the lower pQCD lower limit is employed). We then choose a sufficiently large value for standard deviation $\sigma$ and a reasonable, intermediate value for mean $\mu$ to ensure the prior is still broad (see Table~\ref{tab:prior_choices}). 

If there are no intermediate data, or if they are constant with respect to the mean function in the region $2n_0 \leqslant n_B \leqslant 20n_0-40n_0$, the optimization of the kernel hyperparameters produces a changepoint close to the $\chi$EFT region (e.g., $2n_0 \leqslant n_B \leqslant 5n_0$). As with the stationary kernel case, this happens because of the ``model complexity" term in the log likelihood that favors kernels possessing a small determinant~\cite{JMLR:v24:22-1153}. In this case the hyperparameter optimization will yield a changepoint that allows the stationary kernel with the longer length scale, i.e., $\kappa_{2}$ (pQCD), to persist through the density range of the other kernel, $\kappa_{1}$ ($\chi$EFT). As in Sec.~\ref{sec:stationaryGPs}, the truncated normal hyperprior on the changepoint and width hyperparameters again prevents the changepoint from allowing such a persistence of the correlation structure in pQCD.

We consider the two different mean functions for the GP prior on $F$ that we employed in Sec.~\ref{sec:stationaryGPs}: the zero mean function and the cubic spline. Other forms of the mean function could also be implemented; we use the cubic spline for its guaranteed continuity in its first two derivatives, which allows us to directly compute the corresponding speed of sound squared. We note that exogenous data in the intermediate-density region reduces the influence of the choice of mean function on the conditional (trained) GP's mean. The GP adjusts to accommodate any intermediate-region data, according to Eq.~\eqref{eq:evaluationmean}, and the mean function then has less influence on the final GP $y$. 

\begin{figure*}[t]
    \centering
    \includegraphics[width=\textwidth]{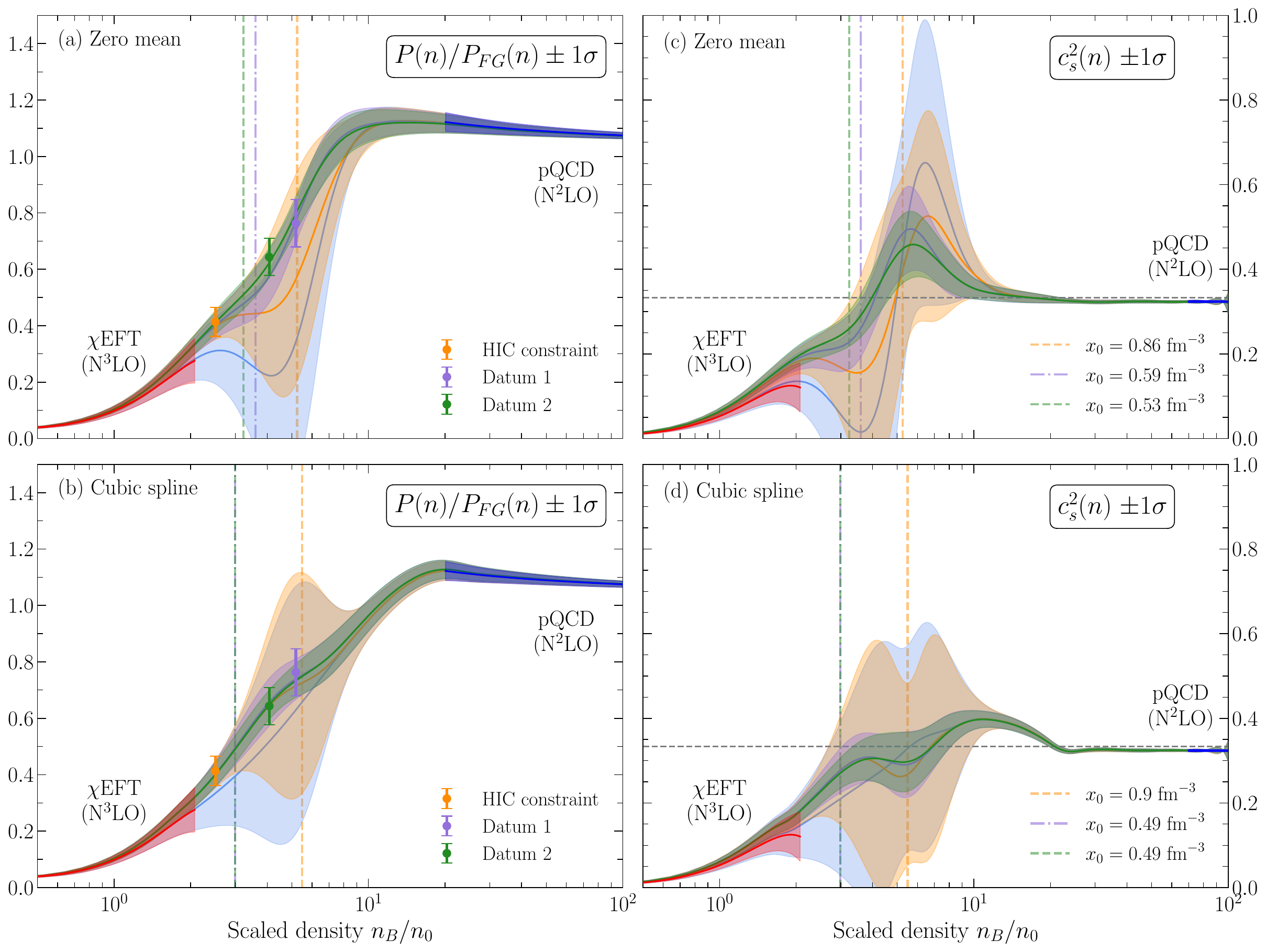}
    \caption{[(a), (b)]: The mixed-model EOS in scaled pressure $P(n_B)/P_{FG}(n_B)$ of neutron star matter with respect to the scaled baryon number density $n_B/n_0$. Here we take a pQCD lower limit of $20n_0$. We employ the changepoint kernel with a sigmoid mixing function and both (a) the zero mean function, and (b) the cubic spline mean function. The color of the data points corresponds to the mixed-model EOS resulting from the addition of that mock constraint. [(c), (d)]: The corresponding speed of sound squared, $c_{s}^{2}(n_B)$, for each iterative addition of the HIC constraint~\cite{Huth:2021bsp} and exogenous theory/experimental data. The locations of the changepoint $x_{0}$ for each iterative training step are shown via the vertical dashed and dash-dot lines, with the color corresponding to each successive iteration.}
    \label{fig:comparison_spline_20n0}
\end{figure*}


\subsection{Inclusion of exogenous data} \label{sec:mockdata}
 
Without data in the intermediate region between $\chi$EFT and pQCD that is gleaned from either model calculations or from novel experimental constraints, the changepoint location is completely driven by the prior, and we cannot investigate how further data would alter our constraints in the mass-radius relationship of a neutron star. In what follows we therefore assume we have additional, exogenous,
information from, e.g., heavy-ion collisions (HIC)~\cite{Sorensen:2023zkk} translated from symmetric into asymmetric matter~\cite{Yao:2023yda}. 
Such information could also come from sophisticated intermediate-density theories that describe QCD dynamics non-perturbatively and possess a full probability distribution. Theories that could satisfy this requirement in the future include Functional Renormalization Group (FRG) methods~\cite{Braun:2022olp, Leonhardt:2019fua} and, potentially, lattice QCD simulations at a finite chemical potential~\cite{Abbott:2024vhj}.

We consider the influence of both experimental and theoretical information as exogenous data. We select a potential HIC constraint by extracting the $2.5n_0$ point and the 95\% uncertainty interval associated with it from the HIC contour in Ref.~\cite{Huth:2021bsp}. We add this ``datum" to our training set for our changepoint GP (see Fig.~\ref{fig:comparison_spline_20n0}a,b). However, we do not refit the mean function to this datum, instead letting the conditional GP adjust the mean to accommodate it, if necessary. 
    
To test the addition of more exogenous data at intermediate densities, we implement a ``greedy" approach, iteratively selecting where to place a potential draw from a theory by finding the baryon number density where the largest uncertainty in scaled pressure is obtained. We then place a point there that is selected from a representation of potential data near the mean curve of the GP, and which possesses conservative errors, $\gtrsim 10\%$ of the mean value (see Fig.~\ref{fig:comparison_spline_20n0}). To create this data set, we use a modified version of the results in Ref.~\cite{Leonhardt:2019fua} at intermediate densities. We assume the exogenous data points we employ in our study have uncorrelated uncertainties. We emphasize that an infinite number of possible data configurations could be chosen for this investigation. Our choices are not meant to exactly capture any existing model, but instead to serve as a demonstration of how our framework can assimilate a set of data, for which we are agnostic as to its origin, into an EOS that already incorporates constraints from $\chi$EFT and pQCD. 


\subsection{EOS Results} \label{sec:changepointresults}

    \begin{figure}
        \centering
        \includegraphics[width=0.91\columnwidth]{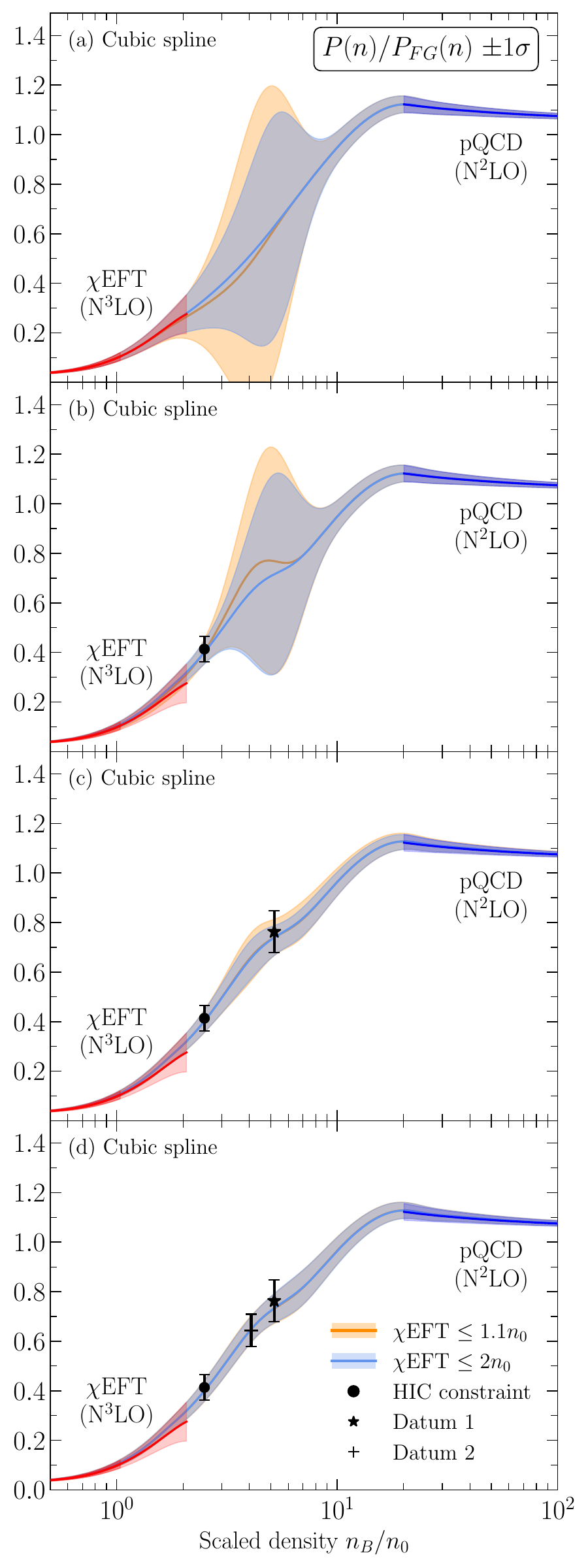}
        \caption{The mixed-model EOS, using a cubic spline mean function, for two chiral cutoff choices: $\leqslant 1.1n_0$ (orange) and $\leqslant 2n_0$ (blue). From the top to the bottom, in each panel we successively add (a) no exogenous data; (b) the HIC constraint~\cite{Huth:2021bsp}; (c) one iteratively chosen data point at $\approx 5n_0$; (d) another iteratively selected point at $\approx 4n_0$. We note a negligible difference between the two EOSs in this final case.}
        \label{fig:chiral_comparison_4panel}
    \end{figure}

We use the changepoint kernel~\eqref{eq:changepointkernel} with the stationary GPs discussed in Sec.~\ref{sec:changepointformalism} as inputs to construct the mixed model EOS for a pQCD lower limit of $20n_0$. We also employ the sigmoid mixing function~\eqref{eq:sigmoid} in this kernel.
We choose our hyperpriors on changepoint $x_{0}$ and width $w$ as discussed in Table~\ref{tab:prior_choices}, and then fit the hyperparameters of the changepoint kernel so that we can predict the EOS at new points across the entire density space. 
    
Figure~\ref{fig:comparison_spline_20n0} presents the resulting mixed-model EOS. We iteratively add the HIC constraint and ``greedy" theory predictions to the training of the GP, producing each mean curve and 68\% uncertainty band shown. We notice that the iterative inclusion of the data causes a general shift in the changepoint location to densities lower than the initial, prior-driven placement. In the case of the zero mean function (Fig.~\ref{fig:comparison_spline_20n0}a,c), the changepoint continues to shift down towards $\chi$EFT as we add each data point. However, in the case of the spline mean function (Fig.~\ref{fig:comparison_spline_20n0}b,d), the changepoint does not change from its location at $3n_0$ after the first exogenous data point is added. This we attribute to a lack of new information; since the spline mean function has already captured the general trend of the EOS, adding more data that causes a perturbative variation of the GP around the location of the mean function will only slightly alter the location of the changepoint. This can be seen directly in Eqs.~\eqref{eq:evaluationmean} and~\eqref{eq:evaluationcov}; the GP mean will be slightly affected by the value of the data around the mean function. The covariance, however, has no dependence on the mean of the training data, only the uncertainties. Hence, only the mean value is affected by the location of the exogenous data point, and a small deviation in the mean of a new point will not greatly shift the overall GP mean.

The stability ($c_{s}^{2}(n_B) > 0$) and causality ($c_{s}^{2}(n_B) \leqslant 1$) of an EOS can again be used here to assess its validity and cull EOS priors (i.e., GP kernels $\kappa_f$) that tend to lead to unphysical EOSs. Since we are constructing our EOS for the scaled pressure, $P(n_B)/P_{FG}(n_B)$, we do not directly impose these conditions, but we post-select among mixed-model EOSs with different GP priors according to whether they obey them or not. For example, when no exogenous data is included, the spline mean function (Fig.~\ref{fig:comparison_spline_20n0}b,d) causes a slightly unstable EOS. However, the GP prior with mean function zero is even more likely to yield an unstable EOS (see Fig.~\ref{fig:comparison_spline_20n0}a,c). The non-zero mean in the GP prior also prevents the appearance of acausal EOSs in the pdf, unlike in Fig.~\ref{fig:comparison_spline_20n0}c. Relatedly, the mixed-model EOSs derived with a non-zero mean in the prior also have scaled pressures and speeds of sound that change more gradually with density. Even after exogenous data are iteratively included, the zero mean function EOS still rises more rapidly and maintains more of the shape of the EOS from pQCD around $10n_0$ than does that for the spline mean function. Thus henceforth we show only results with the spline mean function, as it produces more stable and causal EOSs.

Regardless of the choice of mean function, as the constraints are added in the intermediate region, the EOS uncertainties noticeably shrink and become approximately the size of the data uncertainties once two to three exogenous data are added. This narrowing of the uncertainties stems from the two exogenous model predictions in the intermediate region, i.e., $n_B > 3n_0$. And a noticeable reduction in the error bands of the mixed-model EOS around $n_0$ to $2n_0$ is evident in both Fig.~\ref{fig:comparison_spline_20n0}a and Fig.~\ref{fig:comparison_spline_20n0}b as soon as the HIC constraint is added. The inclusion of this data point shifts the mean curve upward and reduces the mixed-model uncertainties to only encompass the upper 68\% of the $\chi$EFT error band. According to the changepoint locations in Fig.~\ref{fig:comparison_spline_20n0}c and Fig.~\ref{fig:comparison_spline_20n0}d, this region is still heavily weighted in favor of the $\chi$EFT stationary kernel, hence we can rule out any extreme influence of the pQCD covariance function. Instead we attribute this result to the location of the HIC constraint and its small error bars. This has also been seen in another study of the EOS in this region~\cite{Alford:2022bpp}.

Finally, we investigate the choice of $\chi$EFT upper limit density in Fig.~\ref{fig:chiral_comparison_4panel}. We take two different values for comparison: $1.1n_{0}$ and $2n_{0}$, both of which are commonly used as upper limits on $\chi$EFT calculations in the literature~\cite{Drischler:2020hwi, Hebeler:2013nza}. Figure~\ref{fig:chiral_comparison_4panel} successively adds each exogenous data point to the mixed model in each panel, comparing the two EOSs resulting from the different $\chi$EFT upper limits. We let the greedy algorithm choose the intermediate density training set for the case of $\chi$EFT $\leqslant 2n_0$, and we use the same training set in the mixed-model calculation with $\chi$EFT input only at $n_B \leqslant 1.1n_0$. As the data is iteratively added, the differences between the mixed-model EOSs decrease (Fig.~\ref{fig:chiral_comparison_4panel}b,c), until, when all three intermediate-density points are included, they are nearly indistinguishable (Fig.~\ref{fig:chiral_comparison_4panel}d). This method to construct the EOS shows that with sufficient data in the intermediate region, using $\chi$EFT up to $2n_0$ does not cause a different result to emerge than if one only uses $\chi$EFT information up to $n_B \approx n_0$.


\subsection{\texorpdfstring{$M-R$}{M-R} constraints} \label{eq:mrchangepointresults}

\begin{figure}
        \centering
        \includegraphics[width=\columnwidth]{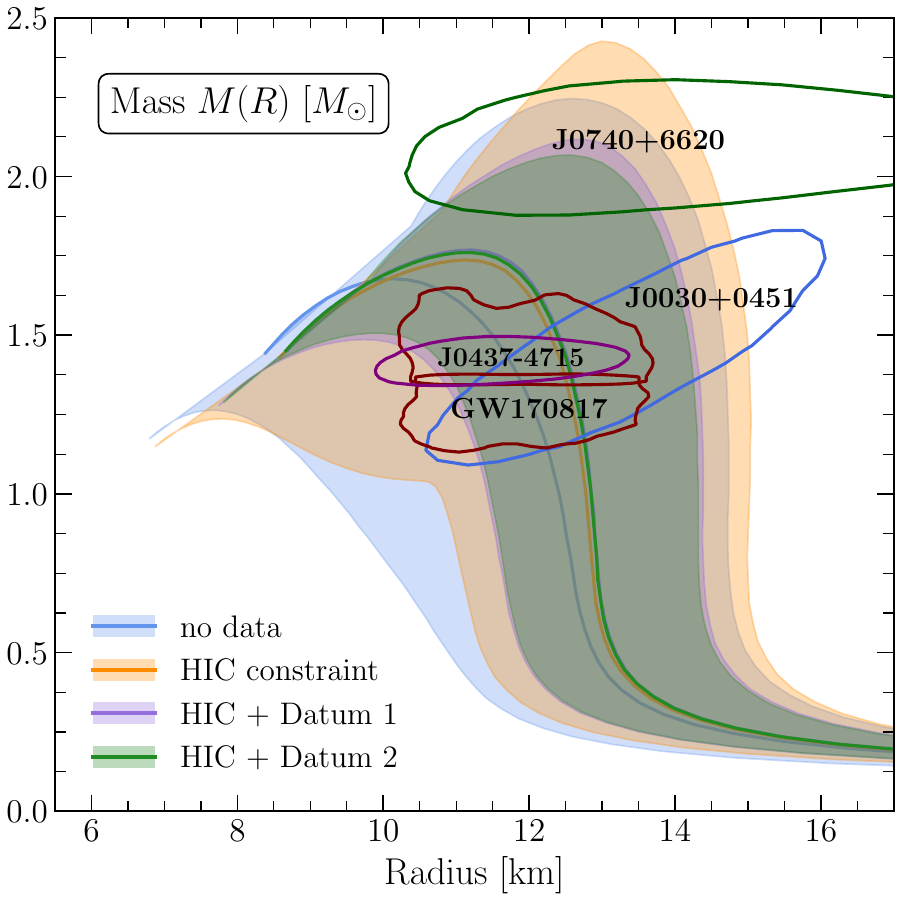}
        \caption{$M-R$ curves at the $1\sigma$ uncertainty level for the changepoint kernel, employing a cubic spline mean function and a sigmoid mixing function. We note that adding the heavy-ion constraint makes the most difference in the shifting of the mean curve upward to higher masses and larger radii. It also reduces the corresponding central density from $8.4n_0$ to $6.9n_0$. Adding constraints from other potential intermediate density measurements or models leads to tighter 68\% uncertainties in the $M-R$ plane. These data have a very small effect on the shape and location of the mean curve, however, and an equally small effect on the corresponding central density (shifting it down from $6.9n_0$ to $6.7n_0$).}
        \label{fig:mr_comparison_spline_20n0_HIC}
    \end{figure}

The overall stiffness of the EOS in the scaled pressure is driven by the HIC constraint at $\approx 2.5n_0$, which can be seen more directly in Fig.~\ref{fig:mr_comparison_spline_20n0_HIC}. This additional constraint causes the mean maximum mass curve to shift upward (from $\approx 1.6 M_{\odot}$ to $\approx 1.7 M_{\odot}$) and to larger radii (from $\approx 10$~km to $\approx 12$~km). Adding one more exogenous data point from the greedy criterion narrows the uncertainties around the mean curve and slightly shifts the mean maximum mass upward. But, it has a much smaller effect than adding the HIC constraint. Adding two points produces a negligible effect on the mean maximum mass, and only a slight tightening of the 68\% uncertainty results. We conclude that once this much data has been added, new data points that occur close to the same mean function (and have similar error bars to those already included) do not significantly affect the EOS constraints. Data with smaller uncertainties would be necessary to further constrain this $M-R$ result.


\section{Summary and outlook} \label{sec:summary}

We extended our BMM framework, first presented in Ref.~\cite{Semposki:2024vnp}, to asymmetric dense matter of neutron stars. For the low-density ($\leqslant 2n_0$) theory input, we constructed the zero-temperature, charge-neutral, $\beta$-equilibrated EOS from $\chi$EFT, including the covariance matrix of uncertainties across densities, using the standard quadratic expansion of the EOS in isospin asymmetry. For the high-density ($\geqslant 20n_0$) theory input, we performed UQ on the cold, $\beta$-equilibrated EOS from pQCD using the BUQEYE truncation error method. We then performed BMM using GPs across the full density input space ($n_B \leqslant 100n_0$) using three classes of stationary GP kernels. We discovered that the RBF kernel still performed very well for neutron star matter and produced realistic $M-R$ results when compared to state-of-the-art NICER and LIGO-Virgo observations. The Mat\'ern 5/2 kernel was the next best choice, but required a non-zero mean function in order to ensure the mixed-model EOS is stable.

To investigate the effect of using a non-stationary kernel on the mixed-model EOS, we used a density-dependent, changepoint kernel with a sigmoid mixing function. This is a more flexible prior on the underlying theory. To test the addition of putative constraints in the intermediate region from potential heavy-ion collision measurements and theory predictions, we iteratively added exogenous data to the mixed-model training set. This was carried out via a greedy approach that placed these data at locations where the mixed-model EOS possessed the largest uncertainties from the previous calculation step. Inclusion of a HIC constraint at $2.5n_0$, derived from Ref.~\cite{Huth:2021bsp}, leads to a large reduction in uncertainty and a markedly stiffer mixed-model EOS. Additional data at $\approx 4n_0$ and $\approx 5n_0$ then serve primarily to tighten the 68\% uncertainty band in both the scaled pressure and $M-R$ space. This points to the need for more precise measurements in the region $n_0 \leqslant n_B \leqslant 5n_0$, where the influence of data on both the mass and radius of neutron stars is substantial~\cite{Drischler:2021bup}. For example, the proposed FRIB400~\cite{FRIB400} will further tighten constraints around $2n_0$. Such data could be used in this BMM framework to improve uncertainty estimates in the interior of neutron stars.

In light of these future advances, it is important to point out that the influence of a specific prior---stationary or non-stationary---in our GP-based approach is only negligible if we have precise astrophysical or heavy-ion data (see Sec.~\ref{sec:changepointresults}), or if we possess enough data that the likelihood becomes the dominant contribution in the mixed model $P(n_B)/P_{FG}(n_B)$ posterior. Our results when adding iterative data to the non-stationary EOS show this: when the heavy-ion collisions data and the first exogenous model data point is included (Datum 1), the EOS is already constrained considerably, and adding Datum 2 only serves to slightly shrink the error band further.

This highlights that the framework developed here is designed for data assimilation. In this respect---and in the use of a GP as a non-parametric extension of a low-density EOS---it is similar to the Relativistic Mean Field Theory + GP approach of Legred et al.~\cite{Legred:2025aar}. However, it is unlike that approach in that the uncertainties of the nuclear-theory input to the framework are quantified, and because the EOS is anchored at asymptotic densities by the pQCD EOS. Exogenous data, regardless of its origin, can be straightforwardly added to the EOS inference, thereby revealing its impact on the scaled pressure, the speed of sound, and on mass and radius predictions. 

On the other hand, Komoltsev et al.~\cite{Komoltsev:2021jzg} developed a way to investigate the influence of pQCD since they use it, together with the maximum speed of sound and thermodynamic conditions on density, to constrain the EOS at lower densities. This interpolation-independent method builds in causality and stability directly. Our results fall within the region in the $P-\varepsilon$ plane identified by Komoltsev et al., especially when we post-select EOS draws for these two conditions.

In contrast, Gorda et al.~\cite{Gorda:2022jvk} build a posterior for the EOS through a likelihood built from astrophysical constraints: X-ray measurements of pulsars, mass measurements from pulsar timing, tidal deformability measurements from GW170817, and the probability of the GW170817 merger remnant being a black hole~\cite{Gorda:2022jvk}. Meanwhile, their GP prior extrapolates the $\chi$EFT EOS up to neutron-star densities. Our mixed-model EOS, which was not informed by these
astrophysical constraints, has a pdf which significantly overlaps the EOS pdfs shown by Gorda in Ref.~\cite{Gorda:2022jvk}.

Our present work can also be used as input to large-scale Bayesian workflows that analyze observations of neutron stars.  Such workflows employ a prior probability distribution for the EOS which is then updated according to astrophysical observations; our framework can provide them with an input distribution derived from quantified theoretical uncertainties. 

Alternatively, our
framework, especially when combined with sophisticated emulators for the computationally expensive TOV equations~\cite{Koehn:2024set,Lalit:2024vmu,Sun:2024nye, Reed:2024urq}, can be used in solving the inverse problem of inferring properties of dense matter from mass-radius constraints. In this way new results from, e.g., future direct gravitational wave detections by the LIGO-Virgo-KAGRA scientific collaborations~\cite{Pang:2022rzc, KAGRA:2021duu, LIGOScientific:2017vwq} as well as 
neutron-star observations by NICER~\cite{Riley:2019yda, Miller:2021qha} and from, e.g., the planned STROBE-X, eXTP, and ATHENA+ campaigns~\cite{STROBE-XScienceWorkingGroup:2019cyd, eXTP:2016rzs, HotEnergeticUniverse2013}, can 
be added to this inference framework or used as data to validate against (as in  Figs.~\ref{fig:mr_comparison_stationary},~\ref{fig:mr_comparison_spline_20n0_HIC}). 

Further planned extensions of our framework include its expansion to multi-dimensional mixing in, e.g., baryon number density and temperature, which is a crucial step to provide a unified EOS for use in neutron-star merger simulations~\cite{Radice:2018pdn}, and potential unification with other modular-based collaborative software packages~\cite{MUSES:2023hyz, ReinkePelicer:2025vuh,Raaijmakers:2025hbz}. 
For this endeavor, improving the modeling of the isospin dependence of the low-density EOS beyond the standard quadratic expansion, modeling phase transitions~\cite{Mroczek:2023zxo,Constantinou:2023ged}, e.g., with compound kernels, such as the changepoint kernels studied here, and identifying where and why \ChiEFT\ breaks down in medium~\cite{Essick:2020flb,Drischler:2021bup,Neill:2024qgg}, will be important, as will the inclusion of uncertainties on the proton fraction of neutron star matter. We note that imposing causality and stability conditions is straightforward if the output variable of the mixing is chosen to be the speed of sound. However, the formulation of a GP prior for the speed of sound is subtler than it is for pressure~\cite{Mroczek:2023zxo}. Future applications of our equation-of-state model-mixing framework could include the speed of sound as an alternative choice for the output space. We provide the community with our full code suite for further use and adaptation via our GitHub repository~\cite{EOS_BMM_ANM}.


\begin{acknowledgments}
We thank Jordan Melendez for insightful suggestions regarding the training of Gaussian processes and for invaluable comments on the manuscript. We also thank Tyler Gorda for sharing the EOS samples from Figure~2 in Ref.~\cite{Gorda:2022jvk}.
A.C.S. acknowledges Andrius Burnelis, Grace Eichler, Jane Kim, Sudhanva Lalit, Yoon Gyu Lee, Joshua M. Maldonado and Bradley McClung for engaging discussions and the Facility for Rare Isotope Beams for hospitality during the completion of this work. C.D. thanks the National Science Foundation’s Physics Frontier Center ``The Network
for Neutrinos, Nuclear Astrophysics, and Symmetries''
(N3AS) for encouragement.
This research was supported by the CSSI program Award OAC-2004601 (BAND collaboration \cite{BAND_Framework}) (A.C.S., R.J.F., D.R.P.), by the U.S. Department of Energy, Office of Science, Nuclear Physics, under Award DE-FG02-93-40756 (A.C.S., D.R.P.), by the FRIB Theory Alliance, under Award DE-SC0013617 (C.D.), by the National Science Foundation Award Nos.\ PHY-2339043 (C.D.) and PHY-2209442 (R.J.F.), by the NUCLEI SciDAC program under award DE-FG02-96ER40963 (R.J.F.), by the Swedish Research Council via a Tage Erlander visiting Professorship, Grant No. 2022-00215 (D.R.P.), and by the US National Science Foundation via Grant PHY-2020275, N3AS PFC (D.R.P.). 
R.J.F.\ also acknowledges support from the ExtreMe Matter Institute EMMI
at the GSI Helmholtzzentrum für Schwerionenforschung GmbH, Darmstadt, Germany.
\end{acknowledgments}


\bibliography{bayesian_refs}


\end{document}